\documentclass[prl,twocolumn]{revtex4-2}
\usepackage{amsmath}
\usepackage{amssymb}
\usepackage{graphicx}
\usepackage{braket}
\usepackage{stmaryrd}
\usepackage{hyperref}
\usepackage{color}

\hypersetup{
    colorlinks,
    citecolor=blue,
    linkcolor=blue,
    urlcolor=blue
}

\newcommand{\sqrbr}[1]{\llbracket #1\rrbracket}

\begin{document}
\title{Irreversibility, Loschmidt Echo, and Thermodynamic Uncertainty Relation}
\author{Yoshihiko Hasegawa}
\email{hasegawa@biom.t.u-tokyo.ac.jp}
\affiliation{Department of Information and Communication Engineering, Graduate
School of Information Science and Technology, The University of Tokyo,
Tokyo 113-8656, Japan}
\date{\today}
\begin{abstract}
Entropy production characterizes irreversibility. This viewpoint allows us to consider the thermodynamic uncertainty relation, which states that a higher precision can be achieved at the cost of higher entropy production, as a relation between precision and irreversibility. 
Considering the original and perturbed dynamics, we show that the precision of an arbitrary counting observable
in continuous measurement of quantum Markov processes is bounded from below by Loschmidt echo
between the two dynamics,
representing the irreversibility of quantum dynamics. 
When considering particular perturbed dynamics, our relation leads to several thermodynamic uncertainty relations,
indicating that our relation provides a unified perspective on classical and quantum thermodynamic uncertainty relations. 
\end{abstract}
\maketitle

\begin{figure*}
\includegraphics[width=16cm]{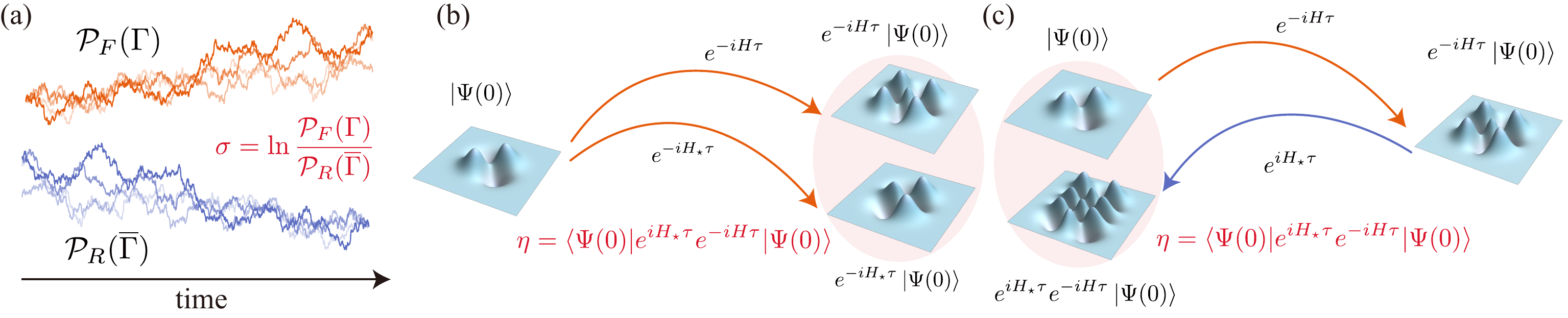} \caption{
Quantification of irreversibility. (a) Entropy production $\sigma$ in classical Markov processes, defined by a log-ratio between $\mathcal{P}_F(\Gamma)$, the probability for observing a trajectory $\Gamma$ in the forward process,
and $\mathcal{P}_R(\overline{\Gamma})$, the probability for observing a time-reversed trajectory $\overline{\Gamma}$ in the reversed process. 
(b) Loschmidt echo $\eta$ in quantum dynamics, which is the fidelity between two states, $e^{-iH\tau} \ket{\Psi(0)}$ and $e^{-iH_\star\tau} \ket{\Psi(0)}$ at $t=\tau$.
These states are obtained through forward time evolution induced by $H$ and $H_\star$, respectively. 
(c) An interpretation of the Loschmidt echo $\eta$ as the fidelity between two states, $\ket{\Psi(0)}$ and $e^{iH_\star\tau}e^{-iH\tau}  \ket{\Psi(0)}$, at $t=0$. 
The latter state is obtained through forward time evolution by $H$ and the subsequent reversed-time evolution by $H_\star$. 
\label{fig:LocshmidtEcho}}
\end{figure*}

\emph{Introduction.---}Thermodynamic uncertainty relation (TUR) \cite{Barato:2015:UncRel,Gingrich:2016:TUP,Pietzonka:2016:Bound,Horowitz:2017:TUR,Pigolotti:2017:EP,Garrahan:2017:TUR,Dechant:2018:TUR,Barato:2018:PeriodicTUR,Terlizzi:2019:KUR,Hasegawa:2019:CRI,Hasegawa:2019:FTUR,Vu:2019:UTURPRE,Vu:2020:TURProtocolPRE,Dechant:2020:FRIPNAS,Vo:2020:TURCSLPRE,Koyuk:2020:TUR,Dechant:2020:ContReversal}
(see \cite{Horowitz:2019:TURReview} for a review) gives a universal relation between precision and thermodynamic cost. It states that $\sqrbr{\jmath}^2 / \braket{\jmath}^2 \ge 2/\braket{\sigma}$, where $\braket{\jmath}$ and $\sqrbr{\jmath}$ are the mean and standard deviation, respectively, of a current observable $\jmath$, and
$\braket{\sigma}$ is the mean of the entropy production.
TUR indicates that a higher precision can be achieved at the cost of higher entropy production. 
The entropy production quantifies the irreversibility of a system. 
Let $\mathcal{P}_F(\Gamma)$ be the probability for observing a trajectory $\Gamma$ in the forward process, and $\mathcal{P}_R (\overline{\Gamma})$ be the probability for observing a time-reversed trajectory $\overline{\Gamma}$ in the reversed process. Then, the entropy production is defined by a log-ratio between $\mathcal{P}_F(\Gamma)$ and $\mathcal{P}_R(\overline{\Gamma})$ [Fig.~\ref{fig:LocshmidtEcho}(a)]:
$\left\langle \sigma\right\rangle =D\left[\mathcal{P}_{F}(\Gamma)||\mathcal{P}_{R}(\overline{\Gamma})\right]\equiv\left\langle \ln\left[\mathcal{P}_{F}(\Gamma)/\mathcal{P}_{R}(\overline{\Gamma})\right]\right\rangle $, where $D[\bullet||\bullet]$ denotes the relative entropy. 
This relation suggests that TUR is a consequence of irreversibility, i.e., the larger the extent of irreversibility, the higher the precision of a thermodynamic machine.

In Newtonian dynamics, despite microscopic reversibility, irreversibility emerges due to the chaotic nature of many-body systems. For chaotic systems, even considering reversed dynamics by reversing the sign of the momenta, an infinitely small perturbation applied to the state yields an exponential divergence from the original reversed dynamics, indicating the infeasibility of such reversed dynamics. 
Thus, the extent of irreversibility can be evaluated through the extent of chaos, which is often quantified by the Lyapunov exponent in classical dynamics. The
Loschmidt echo \cite{Peres:1984:LE,Gorin:2006:LEReview,Goussev:2012:LoschmidtEcho} is an indicator for the effect of small perturbations applied to the Hamiltonian in quantum systems. It can be viewed as a quantum analog of the Lyapunov exponent. 
Consider an isolated quantum system. Given an initial pure state $\ket{\Psi(0)}$, with Hamiltonian $H$ and perturbed Hamiltonian $H_\star$, the Loschmidt echo $\eta$ is defined as follows:
\begin{equation}
\eta \equiv |\braket{\Psi(0)|e^{iH_\star \tau} e^{-iH\tau}|\Psi(0)}|^2.
\label{eq:Loschmidt_def}
\end{equation}Equation~\eqref{eq:Loschmidt_def} evaluates the fidelity between two states, $e^{-iH\tau}\ket{\Psi(0)}$ and $e^{-iH_\star\tau}\ket{\Psi(0)}$, at time $t=\tau$, where $\tau > 0$ [Fig.~\ref{fig:LocshmidtEcho}(b)]. 
These states are generated through the forward time evolution induced by $H$ and $H_\star$, respectively. 
Alternatively, Eq.~\eqref{eq:Loschmidt_def} can be viewed as the fidelity between $\ket{\Psi(0)}$ and $e^{iH_\star \tau} e^{-iH\tau}\ket{\Psi(0)}$ at $t=0$, where the latter state is obtained by applying the forward time evolution by $H$ and 
the subsequent reversed-time evolution by $H_\star$ to $\ket{\Psi(0)}$ [Fig.~\ref{fig:LocshmidtEcho}(c)].
The second interpretation gives a natural extension to classical irreversibility. 
In this Letter, we show that the precision of any counting observable in
the continuous measurement of quantum Markov processes
is bounded from below by the Loschmidt echo.
This relation can be viewed as a quantum extension of classical TURs.
Notably, the obtained quantum TUR holds for any continuous measurement,
which has not been achieved for previous quantum TURs \cite{Erker:2017:QClockTUR,Brandner:2018:Transport,Carollo:2019:QuantumLDP,Liu:2019:QTUR,Guarnieri:2019:QTURPRR,Saryal:2019:TUR,Hasegawa:2020:QTURPRL,Friedman:2020:AtomicTURPRB,Hasegawa:2020:TUROQS,Sacchi:2021:BosonicTUR,Kalaee:2021:QTURPRE,Timpanaro:2021:MostPrecise}. 
When we consider empty dynamics for the perturbed dynamics, the main result appears to be 
reminiscent of a bound obtained in Ref.~\cite{Hasegawa:2020:TUROQS}, whereas the bound in this Letter provides a tighter bound for a classical limit. 
Moreover, when we consider time-scaled perturbed dynamics,
the main result reduces to the bound reported in Ref.~\cite{Hasegawa:2020:QTURPRL},
which covers a classical TUR comprising the dynamical activity \cite{Garrahan:2017:TUR}. 
Our result provides a unified perspective on classical and quantum TURs.

\emph{Results.---}We consider a quantum Markov process described by a Lindblad equation \cite{Lindblad:1976:Generators,Breuer:2002:OpenQuantum}. Let $\rho_S(t)$ be a density operator at time $t$
in the principal system $S$. The time evolution of $\rho_S(t)$ is governed by
\begin{equation}
\dot{\rho}_S = \mathcal{L}\rho_S \equiv  -i\left[H_S,\rho_S\right]+\sum_{m=1}^{M}\mathcal{D}(\rho_S,L_{m}),
\label{eq:Lindblad_def}
\end{equation}where $\dot{\bullet}$ is the time derivative, $\mathcal{L}$ is a Lindblad super-operator, $H_S$ is a Hamiltonian, $\mathcal{D}(\rho_S,L)\equiv L\rho_S L^{\dagger}-\left\{ L^{\dagger}L,\rho_S\right\} /2$
is a dissipator, and $L_{m}$ ($1\le m\le M$ with $M$ being the number of $L_m$) is the $m$th jump operator ($[\bullet,\bullet]$
and $\{\bullet,\bullet\}$ denote the commutator and anticommutator,
respectively). 
Note that $H_S$ is different from the total Hamiltonian $H$, which induces unitary time evolution in the total system. 
For a sufficiently small time interval $\Delta t$,
Eq.~\eqref{eq:Lindblad_def} admits the Kraus representation $\rho_S(t+\Delta t) = \sum_{m=0}^M V_m \rho_S(t) V_m^\dagger$, where 
\begin{align}
V_{0} & \equiv \mathbb{I}_{S}-i\Delta tH_S-\frac{1}{2}\Delta t\sum_{m=1}^{M}L_{m}^{\dagger}L_{m},\label{eq:Jump_V0_def}\\
V_{m} & \equiv\sqrt{\Delta t}L_{m}\;\;\;(1\le m\le M).\label{eq:Jump_Vc_def}
\end{align}Here, $\mathbb{I}_S$ denotes the identity operator in $S$ (the other identity operators are defined similarly). 
$V_0$ corresponds to no jump and $V_m$ ($1 \le m \le M$) to the $m$th jump within the interval $[t,t+\Delta t]$. 
$V_{m}$ ($0 \le m \le M$) satisfies the completeness relation $\sum_{m=0}^{M}V_{m}^{\dagger}V_{m}=\mathbb{I}_{S}$.
$V_m$ defined in Eqs.~\eqref{eq:Jump_V0_def} and \eqref{eq:Jump_Vc_def} are not the only operators consistent with Eq.~\eqref{eq:Lindblad_def}. There are infinitely many operators that can induce the same time evolution.

\begin{figure}
\centering
\includegraphics[width=7cm]{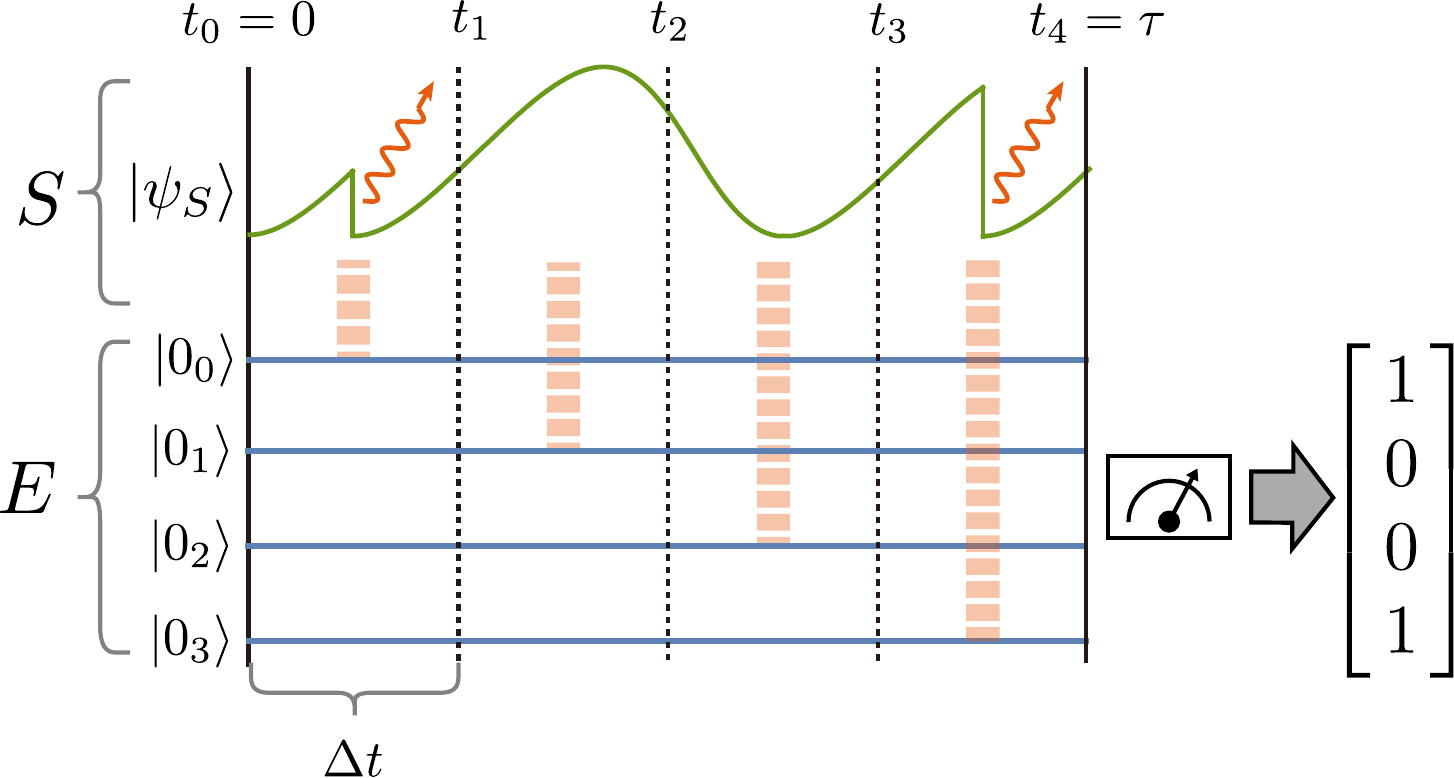} \caption{
Illustration of continuous measurement model for $N=4$. 
The initial states of $S$ and $E$ are $\ket{\psi_S}$ and $\ket{0_3,0_2,0_1,0_0}$, respectively.
The environment subspace $\ket{0_k}$ interacts with $S$ during an interval $[t_k, t_{k+1}]$. Finally, at $t=\tau$,
$E$ is measured with a set of projectors $\{\ket{\boldsymbol{m}}\bra{\boldsymbol{m}}\}_{\boldsymbol{m}}$.
Suppose that the measurement record is $\boldsymbol{m} = [1_3,0_2,0_1,1_0]$.
Then, the state of the principal system undergoes two jump events in the intervals $[t_0,t_1]$ and $[t_3,t_4]$.
\label{fig:simulation}}
\end{figure}

Using the input-output formalism \cite{Guta:2011:AsympNorm,Gammelmark:2014:QCRB,Macieszczak:2016:QMetrology,Gross:2018:ContMeasQubit}, 
employed in studying TURs in a quantum domain \cite{Hasegawa:2020:QTURPRL,Hasegawa:2020:TUROQS}, 
we describe the time evolution generated by the Kraus operators [Eqs.~\eqref{eq:Jump_V0_def} and \eqref{eq:Jump_Vc_def}]
as interactions between the principal system $S$ and environment $E$. 
Let $t=0$ and $t=\tau$ be the initial and final times of time evolution, respectively. 
We discretize the time interval $[0,\tau]$ by dividing it into $N$ intervals, where $N$ is a sufficiently large natural number; in addition, we define $\Delta t \equiv \tau /N$ and $t_k \equiv \Delta t k$ ($t_0=0$ and $t_N=\tau$). 
Here, the orthonormal basis of $E$ is assumed to be $\ket{m_{N-1},...,m_1,m_0}$ ($m_k \in \{0, 1,...,M-1,M\}$), where
a subspace $\ket{m_k}$ interacts with $S$ through a unitary operator $U_{t_{k}}$ during an interval $[t_k, t_{k+1}]$ [Fig.~\ref{fig:simulation}]. 
When the initial states of $S$ and $E$ are $\ket{\psi_S}$ and $\ket{0_{N-1},...,0_1,0_0}$, respectively, 
the composite state at $t=\tau$ is
\begin{align}
\ket{\Psi(\tau)}&=U_{t_{N-1}}\cdots U_{t_{0}}\ket{\psi_S}\otimes\ket{0_{N-1},\cdots,0_{0}}\nonumber\\
&=\sum_{\boldsymbol{m}}V_{m_{N-1}}\cdots V_{m_{0}}\ket{\psi_S}\otimes\ket{m_{N-1},\cdots,m_{0}}.
\label{eq:input_output}
\end{align}Calculating $\mathrm{Tr}_E [\ket{\Psi(\tau)}\bra{\Psi(\tau)}]$ for $\Delta t \to 0$,
we recover the original Lindblad equation of Eq.~\eqref{eq:Lindblad_def}. 
This input-output formalism is referred to as the repeated interaction model in quantum thermodynamics \cite{Horowitz:2013:QJ,Manzano:2018:EPPRX,Santos:2020:JFT},
which was recently used to derive quantum TURs \cite{Miller:2021:QTURQHE,Miller:2021:QTUREP}. Importantly, 
the input-output formalism assumes that the environment is pure so that the 
time-evolved state in $S+E$ is pure [Eq.~\eqref{eq:input_output}], which enables the following calculation. 
Continuous measurement \cite{Wiseman:1996:QJ,Elouard:2018:QTraj} through the environment corresponds to environmental measurement at the final time. 
When we measure the environment at $t=\tau$ through a set of projectors
$\{\ket{\boldsymbol{m}}\bra{\boldsymbol{m}}\}_{\boldsymbol{m}}$ with $\boldsymbol{m} \equiv [m_{N-1},...,m_1,m_0]$,
we obtain a realization of $\boldsymbol{m}$, and the principal system is projected to $V_{m_{N-1}}\cdots V_{m_0} \ket{\psi_S}$ (note that this is unnormalized).
Thus, $\boldsymbol{m}$ comprises a measurement record of continuous measurement. 
Since the evolution of $V_{m_{N-1}}\cdots V_{m_0} \ket{\psi_S}$ is stochastic depending on the measurement record, 
it is referred to as a \emph{quantum trajectory}, which can be described by the stochastic Schr\"odinger equation \cite{Supp:2021:LE}. \nocite{Landi:2018:TextBook}
Figure~\ref{fig:simulation} presents an example of continuous measurement in the input-output formalism for
$M=1$ (i.e., a single jump operator) with $N=4$.
After the measurement with the set of projectors $\{\ket{\boldsymbol{m}}\bra{\boldsymbol{m}}\}_{\boldsymbol{m}}$, suppose we obtain $[1_3,0_2,0_1,1_0]$, where $1$s denote the detection of jumps.
Then, two jump events occurred in the principal system $S$ 
in two intervals $[t_0,t_1]$ and $[t_3,t_4]$.
Let us define the counting observable $\mathcal{C}$
that counts and weights jump events in a quantum trajectory. 
We define $\mathcal{C}$ by an Hermitian operator on $E$, 
which admits the following eigendecomposition:
\begin{align}
\mathcal{C}=\sum_{\boldsymbol{m}}g(\boldsymbol{m})\ket{\boldsymbol{m}}\bra{\boldsymbol{m}}=\sum_{c}c\Upsilon(c),
\label{eq:G_def}
\end{align}where $\Upsilon(c)\equiv\sum_{\boldsymbol{m}:g(\boldsymbol{m})=c}\ket{\boldsymbol{m}}\bra{\boldsymbol{m}}$ and we assume $g(\boldsymbol{0}) = 0$ with $\boldsymbol{0}\equiv [0_{N-1},\ldots,0_1,0_0]$. A set $\{\Upsilon(c)\}_c$ comprises a projection-valued measure. $g(\boldsymbol{m})$ in Eq.~\eqref{eq:G_def} counts and weights jumps in
a measurement record $\boldsymbol{m}$.
The condition $g(\boldsymbol{0}) = 0$ implies that the counting observable should vanish when there are no jump events, which
constitutes a minimum assumption for the counting observable \cite{Hasegawa:2020:TUROQS}. 
For instance, $g(\boldsymbol{m})$ is typically expressed as follows: 
\begin{equation}
g(\boldsymbol{m}) = \sum_{k=0}^{N-1} C_{m_{k}},
\label{eq:gm_def}
\end{equation}where $[C_0,C_1,...,C_M]$ with $C_0=0$ is a real projection vector specifying the weight of each jump (recall $m_k \in \{0,1,\ldots,M-1,M\}$). 
For instance, in Fig.~\ref{fig:simulation} with the weight vector $[C_0, C_1]=[0,1]$, $g(\boldsymbol{m})$ in Eq.~\eqref{eq:gm_def} simply counts the number of jumps in $\boldsymbol{m}$ to yield $g(\boldsymbol{m})=2$ for $\boldsymbol{m} = [1_3,0_2,0_1,1_0]$.
The probability distributions of the counting observable are
$\mathbb{P}(c) \equiv \braket{\Psi | \mathbb{I}_S\otimes \Upsilon(c)|\Psi}$ and $\mathbb{P}_\star(c) \equiv \braket{\Psi_\star | \mathbb{I}_S\otimes \Upsilon(c)|\Psi_\star}$.
The mean and standard deviation are $\braket{\mathcal{C}} \equiv \braket{\Psi(\tau) | \mathbb{I}_S \otimes \mathcal{C} | \Psi(\tau)} = \sum_c c \mathbb{P}(c)$ and 
$\sqrbr{\mathcal{C}}\equiv\sqrt{\braket{\mathcal{C}^{2}}-\braket{\mathcal{C}}^{2}}$, respectively (quantities with a subscript $\star$ should be evaluated for $\ket{\Psi_\star}$ instead  of $\ket{\Psi}$). 

The Lindblad equation of Eq.~\eqref{eq:Lindblad_def} covers classical stochastic processes. 
Consider a classical Markov chain with $N_S$ states.
Such classical states can be represented quantum mechanically by an orthonormal basis
$\{\ket{b_1},\ket{b_2},...,\ket{b_{N_S}} \}$. 
Classical Markov chains can be emulated by setting $H_S = 0$, $L_{ji} = \sqrt{\gamma_{ji}}\ket{b_j}\bra{b_i}$, and
$\rho_S(t) = \sum_i p_i(t) \ket{b_i}\bra{b_i}$, where $\gamma_{ji}$ is a transition rate from $\ket{b_i}$ to $\ket{b_j}$, and $p_i(t)$ is the probability of being $\ket{b_i}$ at time $t$. 
$\mathcal{C}$ with Eq.~\eqref{eq:gm_def} is a reminiscent of the counting observable 
in the classical stochastic thermodynamics, which is defined by $\sum_{j \ne i} \mathsf{C}_{ji}\mathsf{N}_{ji}$ with 
$\mathsf{N}_{ji}$ being the number of transitions from $\ket{b_i}$ to $\ket{b_j}$ in $[0,\tau]$, and $\mathsf{C}_{ji} \in \mathbb{R}$ being its weight. 
The \emph{current} observable, which is an observable of interest in the conventional TUR \cite{Barato:2015:UncRel,Gingrich:2016:TUP}, additionally assumes antisymmetry $\mathsf{C}_{ji} = -\mathsf{C}_{ij}$.

The Loschmidt echo considers the fidelity between the original $\ket{\Psi}$ and perturbed state $\ket{\Psi_\star}$ [Eq.~\eqref{eq:Loschmidt_def}]. 
Let $H_{\star,S}$ and $L_{\star,m}$ ($1 \le m \le M$) be the perturbed Hamiltonian and  jump operators, respectively, in Eqs.~\eqref{eq:Jump_V0_def} and \eqref{eq:Jump_Vc_def}.
We define the Kraus operators of the perturbed dynamics $V_{\star,m}$ by 
Eqs.~\eqref{eq:Jump_V0_def} and \eqref{eq:Jump_Vc_def}, where $H_S$ and $L_m$ should be replaced with $H_{\star,S}$ and $L_{\star,m}$, respectively. Similar to Eq.~\eqref{eq:input_output}, the composite state of the perturbed dynamics at $t=\tau $ is given by
\begin{equation}
\ket{\Psi_{\star}(\tau)}=\sum_{\boldsymbol{m}}V_{\star,m_{N-1}}\cdots V_{\star,m_{0}}\ket{\psi_S}\otimes\ket{m_{N-1},\cdots,m_{0}}.
\label{eq:conj_Vm_def}
\end{equation}Calculating the Loschmidt echo $|\braket{\Psi_\star | \Psi}|^2$ for Eqs.~\eqref{eq:input_output} and \eqref{eq:conj_Vm_def} is not an easy task since the composite state [$\ket{\Psi(\tau)}$ or $\ket{\Psi_\star(\tau)}$], comprising the principal system and environment, is generally inaccessible. 
For continuous measurement, the Loschmidt echo can be explicitly calculated after Refs.~\cite{Gammelmark:2014:QCRB,Molmer:2015:HypoTest}. 
Note that $\braket{\Psi_{\star}(t)|\Psi(t)}=\mathrm{Tr}_{SE}\left[\ket{\Psi(t)}\bra{\Psi_{\star}(t)}\right]=\mathrm{Tr}_{S}\left[\phi(t)\right]$ where $\phi(t) \equiv \mathrm{Tr}_{E}[\ket{\Psi(t)}\bra{\Psi_\star(t)}]$.
Thus, using Eqs.~\eqref{eq:input_output} and \eqref{eq:conj_Vm_def}, $\phi$ satisfies a two-sided Lindblad equation \cite{Gammelmark:2014:QCRB,Molmer:2015:HypoTest}: $\dot{\phi}=\mathcal{K}\phi\equiv-iH_{S}\phi+i\phi H_{{\star},S}+\sum_{m}L_{m}\phi L_{\star,m}^{\dagger}-\frac{1}{2}\sum_{m}[L_{m}^{\dagger}L_{m}\phi+\phi L_{\star,m}^{\dagger}L_{\star,m}]$,
where $\mathcal{K}$ is a super-operator. Note that $\phi$ does not preserve the trace, i.e., $\mathrm{Tr}_S[\phi(t)]\ne 1$ in general. 
By solving the two-sided Lindblad equation, we obtain $\phi(\tau)=e^{\mathcal{K}\tau}\rho_S(0)$, where $\rho_S(0) = \ket{\psi_S}\bra{\psi_S}$ is the initial density operator of the Lindblad dynamics \footnote{Since $\mathcal{K}$ is a super-operator, evaluating $e^{\mathcal{K}\tau}$ requires the calculation in the Liouville space \cite{Supp:2021:LE}}. 
The Loschmidt echo $\eta$ is expressed by $\eta=\left|\mathrm{Tr}_{S}\left[e^{\mathcal{K}\tau}\rho_S(0)\right]\right|^{2}$. 
Importantly, $\eta$ can be specified by quantities of $S$ alone ($H_S$, $L_m$, $H_{\star,S}$, and $L_{\star,m}$). 
Moreover, the Loschmidt echo $|\braket{\Psi_\star|\Psi}|^2$ can be obtained via the consideration of an ancillary qubit \cite{Molmer:2015:HypoTest},
which is a natural extension of the approach used to experimentally measure the Loschmidt echo in closed quantum dynamics. 
The above calculations assumed an initially pure state; however, a generalization to an initially mixed state case is straightforward \cite{Supp:2021:LE}.

Next, we relate the precision of the counting observable $\mathcal{C}$ with the Loschmidt echo $\eta$. 
Let $\mathcal{F}$ be an arbitrary Hermitian operator on $\ket{\Psi}$ and $\ket{\Psi_\star}$. 
$\mathcal{F}$ admits the eigendecomposition $\mathcal{F} = \sum_{z\in \mathcal{Z}} z \Lambda(z)$, where $\mathcal{Z}$ and $\Lambda(z)$
represent a set of distinct eigenvalues of $\mathcal{F}$ and a projector corresponding to $z$, respectively. 
By using the projector $\Lambda(z)$, the fidelity is bounded from above by
\begin{align}
\left|\braket{\Psi_{\star}|\Psi}\right|&\le\sum_{z\in\mathcal{Z}}\left|\braket{\Psi_{\star}|\Lambda(z)|\Psi}\right|\le\sum_{z\in\mathcal{Z}}\sqrt{P(z)}\sqrt{P_{\star}(z)}\nonumber\\&=1-\mathcal{H}^{2}(P,P_{\star}),
\label{eq:upper_bound_overlap}
\end{align}where $P(z)\equiv \braket{\Psi |\Lambda(z) | \Psi}$, $P_\star(z) \equiv \braket{\Psi_\star |\Lambda(z) | \Psi_\star}$, and $\mathcal{H}^2(\bullet, \bullet)$ is the Hellinger distance. 
The first inequality used the triangle inequality, where the equality holds if and only if the direction
of $\braket{\Psi_\star|\Lambda(z)|\Psi}$ in the complex plane is the same for all $z$.
The second inequality used the Cauchy--Schwarz inequality, 
where the equality holds if and only if $\Lambda(z)\ket{\Psi_\star} \propto \Lambda(z)\ket{\Psi}$ for all $z$.
The Hellinger distance is defined as follows:
$\mathcal{H}^2(P,P_\star) =\frac{1}{2} \sum_{z\in\mathcal{Z}} \left( \sqrt{P(z)} - \sqrt{P_\star(z)}  \right)^2$, 
where $0\le \mathcal{H}^2(P,P_\star) \le 1$. 
The Hellinger distance has a lower bound, given the mean and variance \cite{Dashti:2017:Bayes,Katsoulakis:2017:InfoIneq,Nishiyama:2020:HellingerBound}. We use a tighter lower bound recently derived in Ref.~\cite{Nishiyama:2020:HellingerBound,Nishiyama:2020:Entropy} (see Ref.~\cite{Supp:2021:LE} for a brief explanation):
\begin{equation}
\mathcal{H}^{2}(P,P_{\star})\ge1-\left[\left(\frac{\braket{\mathcal{F}}-\braket{\mathcal{F}}_{\star}}{\sqrbr{\mathcal{F}}+\sqrbr{\mathcal{F}}_{\star}}\right)^{2}+1\right]^{-\frac{1}{2}},
\label{eq:Hellinger_bound}
\end{equation}where $\braket{\mathcal{F}}\equiv\braket{\Psi|\mathcal{F}|\Psi}=\sum_{z\in\mathcal{Z}}zP(z)$ and $\sqrbr{\mathcal{F}} \equiv \sqrt{\braket{\mathcal{F}^2} - \braket{\mathcal{F}}^2}$ 
(quantities with a subscript $\star$ should be evaluated for $\ket{\Psi_\star}$ instead  of $\ket{\Psi}$). 
The equality of Eq.~\eqref{eq:Hellinger_bound} holds if and only if $P(z)$ and $P_\star(z)$ are defined on a set consisting of two points. 
Substituting Eq.~\eqref{eq:Hellinger_bound} into Eq.~\eqref{eq:upper_bound_overlap}, we obtain
\begin{equation}
\left(\frac{\sqrbr{\mathcal{F}}+\sqrbr{\mathcal{F}}_{\star}}{\braket{\mathcal{F}}-\braket{\mathcal{F}}_{\star}}\right)^{2}\ge\frac{1}{\eta^{-1}-1}.
\label{eq:main_result}
\end{equation}Note that a similar relation, which is looser than Eq.~\eqref{eq:main_result}, was derived in Ref.~\cite{Holevo:1973:CRI}.
Because $\mathcal{F}$ is arbitrary,
by taking $\mathcal{F}=\mathbb{I}_{S}\otimes\mathcal{C}=\sum_{c}c(\mathbb{I}_{S}\otimes\Upsilon(c))$ in Eq.~\eqref{eq:main_result}, where $\mathcal{C}$ is the counting observable defined in Eq.~\eqref{eq:G_def}, we obtain 
\begin{equation}
\left(\frac{\sqrbr{\mathcal{C}}+\sqrbr{\mathcal{C}}_{\star}}{\braket{\mathcal{C}}-\braket{\mathcal{C}}_{\star}}\right)^{2}\ge\frac{1}{\left|\mathrm{Tr}_{S}\left[e^{\mathcal{K}\tau}\rho_S(0)\right]\right|^{-2}-1},
\label{eq:main_result_TUR}
\end{equation}which is the main result of this Letter. 
The left-hand side of Eq.~\eqref{eq:main_result_TUR}
concerns the counting observable $\mathcal{C}$, whereas
the right-hand side can be calculated through $H_S, H_{S,\star}, L_m, L_{\star,m}$, and
$\rho_S(0)$. 
The left-hand side of Eq.~\eqref{eq:main_result_TUR} can be identified as a similarity between two distributions, $\mathbb{P}(c)$ and $\mathbb{P}_\star(c)$, when $(\braket{\mathcal{C}} - \braket{\mathcal{C}}_\star)^2$ is sufficiently large. 
Equation~\eqref{eq:main_result_TUR} shows that the precision of counting observables improves
when the extent of irreversibility increases, 
which qualitatively agrees with the
classical TURs \cite{Barato:2015:UncRel,Gingrich:2016:TUP}.
Classical TURs have a lower bound based on the entropy production
that characterizes the irreversibility of classical Markov processes. 
Equation~\eqref{eq:main_result_TUR} is reminiscent of a hysteretic TUR \cite{Proesmans:2019:HTURJSM},
which considers an observable of two processes. 
Similarly, Eq.~\eqref{eq:main_result_TUR} includes the mean and variance of two dynamics, the
original and perturbed dynamics. 
The Kraus operators $V_m$ in Eqs.~\eqref{eq:Jump_V0_def} and \eqref{eq:Jump_Vc_def}
are not unique; a different Kraus operator corresponds to a different continuous measurement. We can show that 
Eq.~\eqref{eq:main_result_TUR} still holds for any continuous measurement (any unraveling) \cite{Supp:2021:LE}.
Let us mention the equality condition of Eq.~\eqref{eq:main_result_TUR}. 
The equality of Eq.~\eqref{eq:main_result_TUR} requires the following three conditions:
$\braket{\Psi_\star | \mathbb{I}_S\otimes \Upsilon(c)|\Psi}$ has the same direction in the complex plane for all $c$, $(\mathbb{I}_S\otimes \Upsilon(c))\ket{\Psi_\star} \propto (\mathbb{I}_S\otimes \Upsilon(c)) \ket{\Psi}$ for all $c$,
and $\mathbb{P}(c)$ and $\mathbb{P}_\star(c)$ are defined on a set comprising two points.
For instance, an observable that simply counts the number of jumps satisfies the last condition for sufficiently short $\tau$,
in which the observable takes either $0$ (no jump event) or $1$ (one jump event). 
We perform simulation analysis for Eqs.~\eqref{eq:main_result_TUR} and numerically verify the bounds (see Ref.~\cite{Supp:2021:LE}). 
At this point, the condition $g(\boldsymbol{0}) = 0$ was not used;
however, it hereafter plays an important role when considering particular perturbed dynamics.

We consider a specific case of the empty perturbed dynamics in Eq.~\eqref{eq:main_result_TUR}, i.e., $H_{\star,S} = 0$ and $L_{\star,m} = 0$ for all $m$. 
Then, the composite state of the perturbed dynamics at $t=\tau$ becomes $\ket{\Psi_\star(\tau)} = \ket{\psi_S}\otimes \ket{0_{N-1},\cdots,0_0}$, which is unchanged from the initial state. 
In this case, the Loschmidt echo is $\eta = |\braket{\Psi(0)|\Psi(\tau)}|^2$. 
Since $\mathcal{C}$ in Eq.~\eqref{eq:G_def} assumes $g(\boldsymbol{0}) = 0$, 
$\braket{\mathcal{C}}_\star = 0$ and $\sqrbr{\mathcal{C}}_\star = 0$ for the empty dynamics. 
Equation~\eqref{eq:main_result_TUR} becomes
\begin{equation}
\frac{\sqrbr{\mathcal{C}}^{2}}{\braket{\mathcal{C}}^{2}}\ge\frac{1}{\left|\mathrm{Tr}_{S}\left[e^{-iH_{\mathrm{eff}}\tau}\rho_{S}(0)\right]\right|^{-2}-1},
\label{eq:main_result2}
\end{equation}where $H_\mathrm{eff} \equiv H_S - (i/2)\sum_m L_m^\dagger L_m$ is the effective Hamiltonian (note that $H_\mathrm{eff}$ is non-Hermitian). 
Note that Eq.~\eqref{eq:main_result2} requires continuous measurement that corresponds to Eqs.~\eqref{eq:Jump_V0_def} and \eqref{eq:Jump_Vc_def} \cite{Supp:2021:LE}, whereas Eq.~\eqref{eq:main_result_TUR} holds for an arbitrary continuous measurement. 
The bound of Eq.~\eqref{eq:main_result2} is similar to that obtained in Ref.~\cite{Hasegawa:2020:TUROQS}:
\begin{equation}
\frac{\sqrbr{\mathcal{C}}^{2}}{\braket{\mathcal{C}}^{2}}\ge\frac{1}{\mathrm{Tr}_{S}\left[e^{-iH_{\mathrm{eff}}^{\dagger}\tau}\rho_{S}(0)e^{iH_{\mathrm{eff}}\tau}\right]-1}.
\label{eq:bound_PRL}
\end{equation}In the short time limit $\tau \to 0$, Eqs.~\eqref{eq:main_result2} and \eqref{eq:bound_PRL} reduce to the same bound 
$\sqrbr{\mathcal{C}}^{2}/\braket{\mathcal{C}}^{2}\ge1/\left[\mathrm{Tr}_{S}[\sum_{m}L_{m}^{\dagger}L_{m}\rho_{S}(0)]\tau\right]$, where
the denominator corresponds to the dynamical activity \cite{Maes:2020:FrenesyPR} in classical Markov processes. 
Using the quantum-to-classical mapping explained above, we can obtain a classical limit of Eq.~\eqref{eq:main_result2} as follows:
\begin{equation}
\frac{\sqrbr{\mathcal{C}}^{2}}{\braket{\mathcal{C}}^{2}}\ge\frac{1}{\left(\sum_{i}p_{i}(0)\sqrt{e^{-\tau\sum_{j(\ne i)}\gamma_{ji}}}\right)^{-2}-1}.
\label{eq:main_result_classical_empty}
\end{equation}Similarly, the classical limit of Eq.~\eqref{eq:bound_PRL} is
\begin{equation}
\frac{\sqrbr{\mathcal{C}}^{2}}{\braket{\mathcal{C}}^{2}}\ge\frac{1}{\sum_{i}e^{\tau\sum_{j(\ne i)}\gamma_{ji}}p_{i}(0)-1},
\label{eq:bound_PRL_classical}
\end{equation}where $e^{-\tau\sum_{j(\ne i)}\gamma_{ji}}$ in Eqs.~\eqref{eq:main_result_classical_empty} and \eqref{eq:bound_PRL_classical}
corresponds to the probability of no jump within $[0,\tau]$ starting from $\ket{b_i}$, which
is an experimentally measurable quantity. By using Jensen's inequality, we can show that Eq.~\eqref{eq:main_result_classical_empty} provides a tighter lower bound than Eq.~\eqref{eq:bound_PRL_classical}. 
Contrariwise, when the system approaches the closed quantum dynamics,
the lower bound of Eq.~\eqref{eq:bound_PRL} becomes tighter than that of Eq.~\eqref{eq:main_result2}
(please see Ref.~\cite{Supp:2021:LE} for details).

So far, we have considered the empty perturbed dynamics. 
We now consider a different perturbed dynamics in Eq.~\eqref{eq:main_result_TUR}, a time-scaled perturbed dynamics.
This case is specified by $H_{\star,S} = (1+\varepsilon)H_S$ and $L_{\star,m} = \sqrt{1+\varepsilon}L_m$ for all $m$, where $\varepsilon \in \mathbb{R}$ is an infinitesimally small parameter.
The Lindblad equation of Eq.~\eqref{eq:Lindblad_def} for the perturbed dynamics
becomes $\dot{\rho}_S = (1+\varepsilon)\mathcal{L}\rho_S$, which is identical to the original dynamics, except for its time scale. 
Assume that this Lindblad equation converges to a single steady-state density operator.
Moreover, 
we perform a continuous measurement corresponding to Eqs.~\eqref{eq:Jump_V0_def} and \eqref{eq:Jump_Vc_def}
assuming the condition of Eq.~\eqref{eq:gm_def} for the counting observable.
Then, for $\tau \to \infty$, according to Ref.~\cite{Burgarth:2015:QEst}, the mean and variance of the perturbed dynamics
satisfy $\braket{\mathcal{C}}_\star = (1 + \varepsilon) \braket{\mathcal{C}}$ and $\sqrbr{\mathcal{C}}^2_\star = (1+\varepsilon) \sqrbr{\mathcal{C}}^2$ \cite{Supp:2021:LE}.
This scaling relation is intuitive; since the time scale of the perturbed dynamics is $1+\varepsilon$ times faster (when $\varepsilon > 0$),
jump events occur $1+\varepsilon$ times more frequently within the fixed time interval $[0,\tau]$ for $\tau \to \infty$. 
However, this scaling relation does not necessarily hold with a different continuous measurement.
The left-hand side of Eq.~\eqref{eq:main_result_TUR} becomes $\left(1+\sqrt{1+\varepsilon}\right)^{2}\sqrbr{\mathcal{C}}^{2}/\left(\varepsilon^{2}\braket{\mathcal{C}}^{2}\right)$. 
Since $\ket{\Psi_\star(\tau)}$ depends on $\varepsilon$, we may write $\ket{\Psi_\star(\tau)} = \ket{\Psi(\tau;\varepsilon)}$,
where $\ket{\Psi(\tau)} = \ket{\Psi(\tau;\varepsilon = 0)}$ because $\varepsilon = 0$ case reduces to the original dynamics.
Next, we evaluate the right-hand side of Eq.~\eqref{eq:main_result_TUR}. The fidelity and quantum Fisher information  are related via \cite{Braunstein:1994:QFI}
\begin{equation}
\mathcal{J}(\tau)=\frac{8}{\varepsilon^{2}}\left[1-\left|\braket{\Psi(\tau;\varepsilon)|\Psi(\tau;0)}\right|\right]\,\,\,(\varepsilon\to0),
\label{eq:QFI_fidelity}
\end{equation}where $\mathcal{J}(\tau)$ denotes the quantum Fisher information \cite{Helstrom:1976:QuantumEst,Hotta:2004:QEstimation,Paris:2009:QFI,Liu:2019:QFisherReviewJPA}. 
Substituting the scaling relation of the left-hand side and Eq.~\eqref{eq:QFI_fidelity} into Eq.~\eqref{eq:main_result_TUR}, for $\varepsilon \to 0$, we obtain
\begin{equation}
    \frac{\sqrbr{\mathcal{C}}^{2}}{\braket{\mathcal{C}}^{2}}\ge\frac{1}{\mathcal{J}(\tau)}\,\,\,\,\,(\tau\to\infty).
    \label{eq:QTUR_CM}
\end{equation}Equation~\eqref{eq:QTUR_CM} rederives a quantum TUR obtained in Ref.~\cite{Hasegawa:2020:QTURPRL} through the quantum Cram\'er--Rao inequality \cite{Helstrom:1976:QuantumEst,Hotta:2004:QEstimation,Paris:2009:QFI,Liu:2019:QFisherReviewJPA}. 
In Ref.~\cite{Hasegawa:2020:QTURPRL}, $\mathcal{J}(\tau)$ was explicitly evaluated and 
shown to reduce to dynamical activity in the classical limit. 
$\mathcal{J}(\tau)$ linearly depends on $\tau$, which contrasts with Eq.~\eqref{eq:main_result2}. 
Therefore, a classical TUR comprising dynamical activity \cite{Garrahan:2017:TUR} can be derived as a particular case of Eq.~\eqref{eq:main_result_TUR}.
Since Eqs.~\eqref{eq:QTUR_CM} and \eqref{eq:main_result2} 
depend on $\tau$ linearly and exponentially, respectively,
Eq.~\eqref{eq:QTUR_CM} is tighter than Eq.~\eqref{eq:main_result2}. 
However, Eq.~\eqref{eq:main_result2} requires fewer assumptions on the dynamics and observable;
Eq.~\eqref{eq:main_result2} holds for arbitrary dynamics and for counting observables satisfying $g(\boldsymbol{0})=0$, whereas
Eq.~\eqref{eq:QTUR_CM} is valid only for steady-state dynamics ($\tau \to \infty$) and for the counting observable, which satisfies the 
additional assumption given in Eq.~\eqref{eq:gm_def}. 
The relative entropy between two nearby probability distributions yields the Fisher information, 
which plays a fundamental role in classical stochastic thermodynamics \cite{Hasegawa:2019:CRI,Dechant:2019:MTUR,Ito:2020:TimeTURPRX,Nicholson:2020:TIUncRel}.
Contrariwise, the quantum relative entropy between two nearby density operators does \textit{not} yield the quantum Fisher information \cite{Nagaoka:2005:QFI},
but the fidelity does [Eq.~\eqref{eq:QFI_fidelity}], which indicates that not the quantum relative entropy, but the Loschmidt echo, 
provides a unified perspective on classical and quantum TURs.

\emph{Conclusion.---}In this Letter, we obtained a relation between the Loschmidt echo and the precision of continuous measurement in quantum Markov processes, which can be viewed as a quantum generalization of classical TURs. 
Since the relations derived in this Letter exploited the advantage of
general quantum bounds, which holds for general Hermitian operators, 
we can obtain other thermodynamic relations for the continuous measurement through our approach. 
Indeed, in our follow-up paper \cite{Hasegawa:2021:FPTTUR},
we obtain a TUR for quantum first passage processes using the same technique. 
Moreover, because Eq.~\eqref{eq:main_result_TUR} shows that the upper bound of the Loschmidt echo $|\braket{\Psi_\star|\Psi}|^2$ can be obtained from the continuous measurement, 
a possible application of Eq.~\eqref{eq:main_result_TUR} is related to thermodynamic inference, 
which is actively studied in classical TURs \cite{Li:2019:EPInference,Manikandan:2019:InferEPPRL,Vu:2020:EPInferPRE,Otsubo:2020:EPInferPRE}. 

\begin{acknowledgments}
This work was supported by the Ministry of Education, Culture, Sports, Science and Technology (MEXT) KAKENHI Grant No.~JP19K12153.
\end{acknowledgments}

\end{document}


\title{Supplementary Material for \\``Irreversibility, Loschmidt Echo, and Thermodynamic Uncertainty Relation''}
\author{Yoshihiko Hasegawa}
\email{hasegawa@biom.t.u-tokyo.ac.jp}
\affiliation{Department of Information and Communication Engineering, Graduate
School of Information Science and Technology, The University of Tokyo,
Tokyo 113-8656, Japan}

\maketitle
This supplementary material describes the calculations introduced in the main text. Equation and figure numbers are prefixed with S (e.g., Eq.~(S1) or Fig.~S1). Numbers without this prefix (e.g., Eq.~(1) or Fig.~1) refer to items in the main text.

\section{Liouville space representation\label{sec:Liouville_space}}
The Lindblad equation describes the time evolution of a density operator $\rho_S$ in a Hilbert space; it can be equivalently described in a Liouville space.
Here, we introduce the Liouville space representation according to Ref.~\cite{Landi:2018:TextBook}.
Given an orthonormal basis $\ket{i}$ in $S$, the density operator is
\begin{equation}
\rho_S = \sum_{i,j} \varrho_{ij}\ket{i}\bra{j}.
\label{eq:rho_def}
\end{equation}We introduce a vectorization of $\rho_S$ by
\begin{equation}
\mathrm{vec}(\rho_S) \equiv \sum_{i,j}\varrho_{ij} \ket{j}\otimes \ket{i}.
\label{eq:rho_vec_def}
\end{equation}Note that $\mathrm{vec}(\rho_S)$ belongs to a Liouville space. Using the vectorization, we can express
the two-sided Lindblad equation in a Liouville space. 
The two-sided Lindblad equation, introduced in the main text, is given by
\begin{equation}
\dot{\phi}=\mathcal{K}\phi\equiv-iH_{S}\phi+i\phi H_{{\star},S}+\sum_{m}L_{m}\phi L_{\star,m}^{\dagger}-\frac{1}{2}\sum_{m}\left[L_{m}^{\dagger}L_{m}\phi+\phi L_{\star,m}^{\dagger}L_{\star,m}\right],
\label{eq:two_sided_LE_def}
\end{equation}where $H_S$ and $L_m$ represent the Hamiltonian and jump operators of the original dynamics, respectively; 
$H_{\star,S}$ and $L_{\star, m}$ represent the Hamiltonian and jump operator in perturbed dynamics. 
We assumed that there are $M$ jump operators $\{L_1,L_2,\ldots,L_M\}$. 
The Liouville representation of Eq.~\eqref{eq:two_sided_LE_def} is expressed as follows:
\begin{equation}
\frac{d}{dt} \mathrm{vec}(\phi) = \hat{\mathcal{K}}\mathrm{vec}(\phi),
\label{eq:twosided_Liouville_def}
\end{equation}where $\hat{\mathcal{K}}$ is defined by
\begin{equation}
\hat{\mathcal{K}}\equiv-i\left[\mathbb{I}_{S}\otimes H_{S}-H_{\star,S}^{\top}\otimes\mathbb{I}_{S}\right]+\sum_{m}\left[L_{\star,m}^{*}\otimes L_{m}-\frac{1}{2}\mathbb{I}_{S}\otimes L_{m}^{\dagger}L_{m}-\frac{1}{2}(L_{\star,m}^{\dagger}L_{\star,m})^{\top}\otimes\mathbb{I}_{S}\right],
\label{eq:Khat_def}
\end{equation}with $*$ and $\top$ denoting the complex conjugate and the transpose, respectively.
Since Eq.~\eqref{eq:twosided_Liouville_def} is a simple linear differential equation, its solution is given by
\begin{equation}
\mathrm{vec}(\phi(\tau))=e^{\hat{\mathcal{K}}\tau}\mathrm{vec}(\rho_S(0)),
\label{eq:phi_solution}
\end{equation}which is abbreviated as $\phi(\tau) = e ^{\mathcal{K}\tau}\rho_S(0)$ in the main text. 

In the main text, we consider a case where the perturbed dynamics is empty, i.e., $H_{\star,S} = 0$ and $L_{\star,m}=0$ for all $m$.
Here, Eq.~\eqref{eq:two_sided_LE_def} becomes
\begin{equation}
\dot{\phi}=\left[-iH_{S}-\frac{1}{2}\sum_{m}L_{m}^{\dagger}L_{m}\right]\phi,
\label{eq:two_sided_empty}
\end{equation}which is a simple linear differential equation. Thus, the solution of Eq.~\eqref{eq:two_sided_empty} can be represented in a Hilbert space as follows:
\begin{equation}
\phi(\tau)=e^{\left[-iH_{S}-\frac{1}{2}\sum_{m}L_{m}^{\dagger}L_{m}\right]\tau}\rho_{S}(0).
\label{eq:phi_solution_empty}
\end{equation}

\section{Mixed state case\label{sec:mixed_state_case}}

In the main text, we consider a two-sided Lindblad equation for initially pure states. 
Here, we consider a two-sided Lindblad equation for an initially mixed state. 

Let $\rho_S$ be the initial mixed state in $S$. We consider an ancilla $A$ that purifies $\rho_S$. 
Let $\ket{\tilde{\psi}_{SA}}$ in $S+A$ be a purification of $\rho_S$:
\begin{equation}
\rho_S = \mathrm{Tr}_A \left[ \ket{\tilde{\psi}_{SA}}\bra{\tilde{\psi}_{SA}} \right].
\label{eq:purification}
\end{equation}We want to define the time evolution on a pure state in $S+A$. 
We define the following Kraus operators on $S+A$:
\begin{equation}
\tilde{V}_{m}\equiv V_{m}\otimes\mathbb{I}_{A}\;\;\;\;\;(0\le m\le M),
\label{eq:Vm_tilde_def}
\end{equation}where $V_m$ is defined in Eqs.~\JumpUVOUdef{} and \JumpUVcUdef{}.
By applying $\tilde{V}_m$ to the purified state $\ket{\tilde{\psi}_{SA}}$ and tracing the ancilla $A$, we obtain
\begin{align}
\mathrm{Tr}_{A}\left[\sum_{m=0}^{M}\tilde{V}_{m}\ket{\tilde{\psi}_{SA}}\bra{\tilde{\psi}_{SA}}\tilde{V}_{m}^{\dagger}\right]&=\mathrm{Tr}_{A}\left[\sum_{m=0}^{M}\left(V_{m}\otimes\mathbb{I}_{A}\right)\ket{\tilde{\psi}_{SA}}\bra{\tilde{\psi}_{SA}}\left(V_{m}^{\dagger}\otimes\mathbb{I}_{A}\right)\right]\nonumber\\&=\sum_{m=0}^{M}V_{m}\mathrm{Tr}_{A}\left[\ket{\tilde{\psi}_{SA}}\bra{\tilde{\psi}_{SA}}\right]V_{m}^{\dagger}\nonumber\\&=\sum_{m=0}^{M}V_{m}\rho_{S}V_{m}^{\dagger},
\label{eq:consistence}
\end{align} indicating that $\tilde{V}_m$ induces a consistent time evolution for $\rho_S$ in $S$. 
Using $\tilde{V}_m$ defined in Eq.~\eqref{eq:Vm_tilde_def}, similar to Eq.~\inputUoutput{}, a pure state in $S+A+E$ at $t=\tau$ is represented by
\begin{equation}
\ket{\tilde{\Psi}(\tau)}=\sum_{\boldsymbol{m}}\tilde{V}_{m_{N-1}}\cdots\tilde{V}_{m_{0}}\ket{\tilde{\psi}_{SA}}\otimes\ket{m_{N-1},\cdots,m_{0}}.
\label{eq:Psi_tilde_def}
\end{equation}As stated in the main text, $\boldsymbol{m} = [m_{N-1},...,m_1,m_0]$ is a measurement record of the observation of the environment $E$ with projector $\{\ket{\boldsymbol{m}}\bra{\boldsymbol{m}}\}_{\boldsymbol{m}}$.
For $\ket{\tilde{\Psi}(\tau)}$, we calculate the probability of obtaining $\boldsymbol{m}$:
\begin{align}
P(m_{N-1},...,m_{0})&=\bra{\tilde{\Psi}(\tau)}\mathbb{I}_{S}\otimes\mathbb{I}_{A}\otimes\ket{m_{N-1},...,m_{0}}\braket{m_{N-1},...,m_{0}|\tilde{\Psi}(\tau)}\nonumber\\&=\braket{\tilde{\psi}_{SA}|\tilde{V}_{m_{0}}^{\dagger}...\tilde{V}_{m_{N-1}}^{\dagger}\tilde{V}_{m_{N-1}}...\tilde{V}_{m_{0}}|\tilde{\psi}_{SA}}\nonumber\\&=\mathrm{Tr}_{SA}\left[\tilde{V}_{m_{N-1}}...\tilde{V}_{m_{0}}\ket{\tilde{\psi}_{SA}}\bra{\tilde{\psi}_{SA}}\tilde{V}_{m_{0}}^{\dagger}...\tilde{V}_{m_{N-1}}^{\dagger}\right]\nonumber\\&=\mathrm{Tr}_{S}\left[V_{m_{N-1}}...V_{m_{0}}\mathrm{Tr}_{A}\left[\ket{\tilde{\psi}_{SA}}\bra{\tilde{\psi}_{SA}}\right]V_{m_{0}}^{\dagger}...V_{m_{N-1}}^{\dagger}\right]\nonumber\\&=\mathrm{Tr}_{S}\left[V_{m_{N-1}}...V_{m_{0}}\rho_{S}V_{m_{0}}^{\dagger}...V_{m_{N-1}}^{\dagger}\right].
\label{eq:Pm_def}
\end{align}Thus, 
statistics of $\boldsymbol{m}$ obtained by quantum trajectories induced by $V_m$ with an initially mixed state $\rho_S$ are identical to the measurement on $\ket{\tilde{\Psi}(\tau)}$ by $\mathbb{I}_S\otimes \mathbb{I}_A \otimes \mathcal{C}$,
where $\mathcal{C}$ is the counting observable defined in Eq.~\GUdef{}.

Similarly, we introduce the Kraus operators $\tilde{V}_{\star,m}$, representing perturbed dynamics:
\begin{equation}
\tilde{V}_{\star,m}\equiv V_{\star,m}\otimes\mathbb{I}_{A}\;\;\;\;\;(0\le m\le M),
\label{eq:V_star_m_def}
\end{equation}where $V_{\star,m}$ is defined in the main text. 
Similar to Eq.~\eqref{eq:Psi_tilde_def}, the pure state of the conjugate dynamics in $S+A+E$ at $t=\tau $ is expressed as follows:
\begin{equation}
\ket{\tilde{\Psi}_{\star}(\tau)}=\sum_{\boldsymbol{m}}\tilde{V}_{\star,m_{N-1}}\cdots\tilde{V}_{\star,m_{0}}\ket{\tilde{\psi}_{SA}}\otimes\ket{m_{N-1},\cdots,m_{0}}.
\label{eq:Psi_tilde_star_def}
\end{equation}As stated in the main text, we can compute the fidelity as follows:
\begin{align}
\braket{\tilde{\Psi}_{\star}(\tau)|\tilde{\Psi}(\tau)}&=\mathrm{Tr}_{SAE}\left[\ket{\tilde{\Psi}(\tau)}\bra{\tilde{\Psi}_{\star}(\tau)}\right]\nonumber\\&=\mathrm{Tr}_{SA}\left[\sum_{\boldsymbol{m}}\tilde{V}_{m_{N-1}}\cdots\tilde{V}_{m_{0}}\ket{\tilde{\psi}_{SA}}\bra{\tilde{\psi}_{SA}}\tilde{V}_{\star,m_{0}}^{\dagger}\cdots\tilde{V}_{\star,m_{N-1}}^{\dagger}\right]\nonumber\\&=\mathrm{Tr}_{S}\left[\sum_{\boldsymbol{m}}V_{m_{N-1}}\cdots V_{m_{0}}\rho_{S}V_{\star,m_{0}}^{\dagger}\cdots V_{\star,m_{N-1}}^{\dagger}\right].
\label{eq:fidelity_calc}
\end{align}The last line of Eq.~\eqref{eq:fidelity_calc} gives a two-sided Lindblad equation with initial state $\rho_S$.

\section{Measurement operator on the environment\label{sec:measurement_op}}

The Kraus operator $V_m$ in Eqs.~\JumpUVOUdef{} and \JumpUVcUdef{} is not unique. 
Let $\mathcal{B}$ be a unitary matrix. Any Kraus operator $Y_m$ defined by
\begin{equation}
Y_{n} = \sum_{m}\mathcal{B}_{nm}V_{m},
\label{eq:Yn_def}
\end{equation}gives the same time evolution as $V_m$, i.e., $\sum_m V_m \rho_S V_m^\dagger = \sum_m Y_m \rho_S Y_m^\dagger$. 
A different Kraus operator corresponds to a different measurement on the environment $E$. 
The basis of the $k$th environmental subspace is $\ket{m_k}$, where $m_k \in \{0,1,...,M\}$.
In this section, we use $\ket{m}$ to denote the subspace $\ket{m_k}$. 
The Kraus operator of Eqs.~\JumpUVOUdef{} and \JumpUVcUdef{}
corresponds to a measurement with basis $\ket{m}$ for each environmental subspace. 
Let us consider the basis $\ket{\alpha}$, representing a basis different from $\ket{m}$. 
$\ket{\alpha}$ is related to $\ket{m}$ through
\begin{equation}
\ket{\alpha} = \sum_m \mathcal{A}_{m\alpha} \ket{m},
\label{eq:phik_def}
\end{equation}where $\mathcal{A}$ is a unitary operator. Direct calculation shows that the unitary operator $\mathcal{A}$ satisfies $\mathcal{A} = \mathcal{B}^\dagger$, 
indicating that the unitary freedom in the Kraus operator corresponds to that in the measurement basis. 
Using $\ket{\alpha}$ and $Y_n$, Eq.~\inputUoutput{} becomes
\begin{align}
\ket{\Psi(\tau)}=\sum_{\boldsymbol{\alpha}}Y_{\alpha_{N-1}}\cdots Y_{\alpha_{0}}\ket{\psi_S}\otimes\ket{\alpha_{N-1},\cdots,\alpha_{0}},
\label{eq:input_output_alpha}
\end{align}where $\boldsymbol{\alpha} \equiv [\alpha_{N-1},...,\alpha_1,\alpha_0]$. In the main text, we consider the counting observable defined in Eq.~\GUdef{} and a measurement on the environment by the set of projectors
$\{\ket{\boldsymbol{m}}\bra{\boldsymbol{m}}\}_{\boldsymbol{m}}$.
Here, we obtain a realization of $\boldsymbol{m}$, and the principal system is projected to $V_{m_{N-1}}\cdots V_{m_0} \ket{\psi}$.
Similarly, we can employ a different set of projectors $\{\ket{\boldsymbol{\alpha}}\bra{\boldsymbol{\alpha}}\}_{\boldsymbol{\alpha}}$ for the environmental measurement. 
Using $\ket{\boldsymbol{\alpha}}\bra{\boldsymbol{\alpha}}$, we obtain a realization of $\boldsymbol{\alpha}$, and the principal system is projected to
$Y_{\alpha_{N-1}}\cdots Y_{\alpha_{0}}\ket{\psi}$. 
We can consider an Hermitian observable on the environment, which is expressed by
\begin{equation}
\tilde{\mathcal{C}}=\sum_{\boldsymbol{\alpha}}\tilde{g}(\boldsymbol{\alpha})\ket{\boldsymbol{\alpha}}\bra{\boldsymbol{\alpha}},
\label{eq:G_def_alpha}
\end{equation}where $\tilde{g}(\boldsymbol{\alpha})$ is an arbitrary function of $\boldsymbol{\alpha}$. Since $\braket{\Psi_\star | \Psi}$ does not depend on
the environmental basis, Eq.~\mainUresultUTUR{} should also hold for $\tilde{\mathcal{C}}$ :
\begin{equation}
\left(\frac{\sqrbr{\tilde{\mathcal{C}}}+\sqrbr{\tilde{\mathcal{C}}}_{\star}}{\braket{\tilde{\mathcal{C}}}-\braket{\tilde{\mathcal{C}}}_{\star}}\right)^{2}\ge\frac{1}{\left|\mathrm{Tr}_{S}\left[e^{\mathcal{K}\tau}\rho_S(0)\right]\right|^{-2}-1}.
\label{eq:main_result_alpha}
\end{equation}$\ket{\alpha}$ can be an arbitrary orthonormal basis in the environment, indicating that the main result of Eq.~\mainUresultUTUR{} holds for any continuous measurement. 

In the main text, we consider the specific case of $H_{\star, S} = 0$ and $L_{\star,m} = 0$ for all $m$. 
Here, the composite state of the conjugate dynamics at $t=\tau$ is $\ket{\Psi_\star(\tau)} = \ket{\psi_{S}}\otimes \ket{0_{N-1},\cdots,0_0}$, which is unchanged from the initial state. The counting observable $\mathcal{C}$ of Eq.~\GUdef{} satisfies
\begin{align}
\braket{\mathcal{C}}_{\star}&=\braket{\Psi_{\star}(\tau)|\mathbb{I}_{S}\otimes\mathcal{C}|\Psi_{\star}(\tau)}=0,\label{eq:G_zero}\\
\braket{\mathcal{C}^{2}}_{\star}&=\braket{\Psi_{\star}(\tau)|\mathbb{I}_{S}\otimes\mathcal{C}^{2}|\Psi_{\star}(\tau)}=0,\label{eq:G2_zero}
\end{align}leading Eq.~\mainUresultUTUR{} to Eq.~\mainUresultII{}. 
For $\tilde{\mathcal{C}}$ defined in Eq.~\eqref{eq:G_def_alpha}, we obtain
\begin{align}
\braket{\tilde{\mathcal{C}}}_{\star}&=\braket{\Psi_{\star}(\tau)|\mathbb{I}_{S}\otimes\tilde{\mathcal{C}}|\Psi_{\star}(\tau)}\ne 0,\label{eq:Gtilde_zero}\\
\braket{\tilde{\mathcal{C}}^{2}}_{\star}&=\braket{\Psi_{\star}(\tau)|\mathbb{I}_{S}\otimes\tilde{\mathcal{C}}^{2}|\Psi_{\star}(\tau)}\ne0.\label{eq:Gtilde2_zero}
\end{align}Equations~\eqref{eq:Gtilde_zero} and \eqref{eq:Gtilde2_zero} show that 
\begin{equation}
\frac{\sqrbr{\tilde{\mathcal{C}}}^{2}}{\braket{\tilde{\mathcal{C}}}^{2}}\ngeq\frac{1}{\left|\mathrm{Tr}_{S}\left[e^{-iH_{\mathrm{eff}}\tau}\rho_{S}(0)\right]\right|^{-2}-1},
\label{eq:main_result2_alpha}
\end{equation}where $H_\mathrm{eff} \equiv H_S - \frac{i}{2}\sum_m L_m^\dagger L_m$. 
Thus, Eq.~\mainUresultII{} depends on how we perform continuous measurement. 

\section{Time-scaled perturbed dynamics\label{sec:scaling}}

\begin{figure}
\centering
\includegraphics[width=10cm]{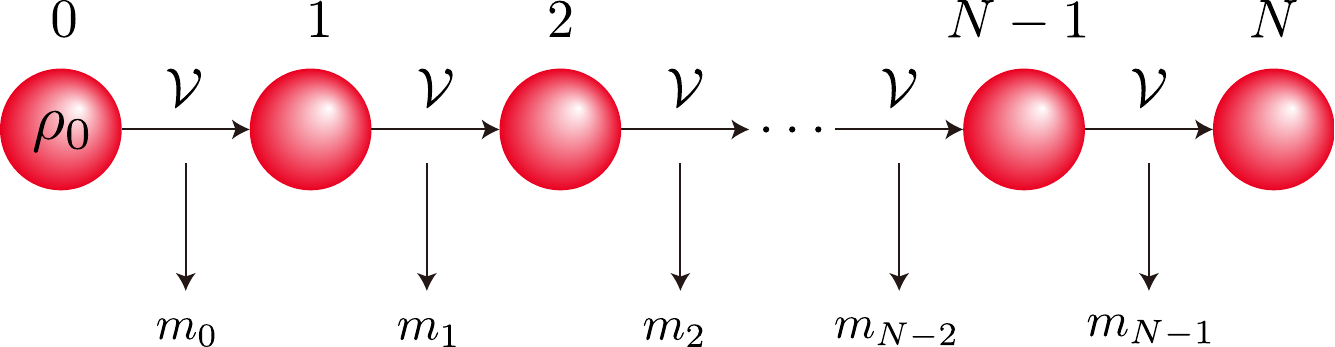} \caption{
Sequential measurement induced by $\mathcal{V}$ at each step. We sequentially apply $V_m$ and record the output $m_i$. 
\label{fig:seq_measurement}}
\end{figure}

In the main text, we consider the following perturbed operators:
\begin{equation}
H_{\star,S} = (1+\varepsilon)H_S,\;\;\;L_{\star,m} = \sqrt{1 + \varepsilon}L_m,
\label{eq:HS_Lm_perturb}
\end{equation}where $\varepsilon \in \mathbb{R}$ is an infinitesimally small parameter. As mentioned in the main text,
the Lindblad equation of Eq.~\eqref{eq:HS_Lm_perturb} becomes $\dot{\rho}_S = (1+\varepsilon)\mathcal{L}\rho_S$ [cf. Eq.~\LindbladUdef{}], 
which is identical to the original dynamics, except for its time scale. 

We will show the following scaling relations from Eq.~\eqref{eq:HS_Lm_perturb}, which hold for $\tau \to \infty$.
\begin{align}
\braket{\mathcal{C}}_\star = (1 + \varepsilon) \braket{\mathcal{C}},\label{eq:mean_scaling_S}\\
\sqrbr{\mathcal{C}}^2_\star = (1+\varepsilon) \sqrbr{\mathcal{C}}^2.\label{eq:var_scaling_S}
\end{align}The Kraus representation is given by
\begin{equation}
\mathcal{V}(\bullet) \equiv \sum_{m=0}^M V_m \bullet V_m^\dagger.
\label{eq:Kraus_V_def_S}
\end{equation}Continuous measurement can be identified as a sequential measurement induced by Eq.~\eqref{eq:Kraus_V_def_S}.
Suppose we sequentially perform a selective measurement $N$ times as shown in Fig.~\ref{fig:seq_measurement}. 
Then, we obtain $\boldsymbol{m} \equiv [m_0,m_1,\ldots,m_{N-1}]$ ($m_k \in \{0,1,\ldots,M\}$), which corresponds to a record of continuous measurement.  
Let us consider the following average:
\begin{equation}
\mathcal{S}_{N}\equiv\frac{1}{N}\sum_{k=0}^{N-1}C_{m_{k}},
\label{eq:sum_def}
\end{equation}where $[C_0,C_1,\ldots,C_M]$ is a projection vector specifying the weight of each jump (see the main text). 
References~\cite{Guta:2011:AsympNorm,Burgarth:2015:QEst} showed that $\mathcal{S}_N$ of the sequential measurement satisfies the following scaling for $N\to \infty$:
\begin{align}
\braket{\mathcal{S}_{N}}=O(N^{0}),\;\;\;\sqrbr{\mathcal{S}_{N}}^{2}=\braket{\mathcal{S}_{N}^{2}}-\braket{\mathcal{S}_{N}}^{2}=O(N^{-1}).
\label{eq:SN_scaling}
\end{align}Relations of Eq.~\eqref{eq:SN_scaling} indicate that, for $N\to \infty$, $\mathcal{S}_N$ scales as if $\boldsymbol{m}$ were independent and identically distributed. 
Indeed, Refs.~\cite{Guta:2011:AsympNorm,Burgarth:2015:QEst} showed that $\mathcal{S}_N$ satisfies the central limit theorem for $N\to \infty$. 
The counting observable, defined in Eq.~\GUdef{} in the main text, is
\begin{equation}
\mathcal{C}=\sum_{\boldsymbol{m}}g(\boldsymbol{m})\ket{\boldsymbol{m}}\bra{\boldsymbol{m}},
\label{eq:C_def_S}
\end{equation}where $g(\boldsymbol{0}) = 0$ with $\boldsymbol{0} \equiv [0_{N-1},\ldots,0_1,0_0]$. 
When considering the time-scaled perturbed dynamics, we specifically consider [Eq.~\gmUdef{}]
\begin{equation}
    g(\boldsymbol{m}) = \sum_{k=0}^{N-1} C_{m_{k}},
    \label{eq:gm_explicit_def_S}
\end{equation}which satisfies $g(\boldsymbol{0}) = 0$.
Because $\tau = N \Delta t$, for $\tau \to \infty$, the counting observable with Eq.~\eqref{eq:gm_explicit_def_S} scales as 
\begin{equation}
\braket{\mathcal{C}}=\Braket{\sum_{k=0}^{N-1}C_{m_{k}}}=O(\tau),\;\;\;\sqrbr{\mathcal{C}}^{2}=\Braket{\left(\sum_{k=0}^{N-1}C_{m_{k}}\right)^{2}}-\braket{\mathcal{C}}^{2}=O(\tau),
\label{eq:C_scaling}
\end{equation}which follows from Eq.~\eqref{eq:SN_scaling}. Note that, for finite $\tau$, Eq.~\eqref{eq:C_scaling} does not hold in general. 

As shown in the main text, the Kraus operators are given by [Eqs.~\JumpUVOUdef{} and \JumpUVcUdef{}]
\begin{align}
V_{0} & = \mathbb{I}_{S}-i\Delta tH_S-\frac{1}{2}\Delta t\sum_{m=1}^{M}L_{m}^{\dagger}L_{m},\label{eq:Jump_V0_def_S}\\
V_{m} & = \sqrt{\Delta t}L_{m}\;\;\;(1\le m\le M).\label{eq:Jump_Vc_def_S}
\end{align}Therefore, the Kraus operators of the perturbed dynamics are
\begin{align}
V_{\star,0}&=\mathbb{I}_{S}-i\Delta tH_{\star,S}-\frac{1}{2}\Delta t\sum_{m=1}^{M}L_{\star,m}^{\dagger}L_{\star,m}=\mathbb{I}_{S}-i\Delta t^{\prime}H_{S}-\frac{1}{2}\Delta t^{\prime}\sum_{m=1}^{M}L_{m}^{\dagger}L_{m},\label{eq:V0_star_def_S}\\V_{\star,m}&=\sqrt{\Delta t}L_{\star,m}=\sqrt{\Delta t^{\prime}}L_{m}\;\;\;(1\le m\le M),\label{eq:Vm_star_def_S}
\end{align}where we defined $t^\prime \equiv (1+\varepsilon)t$. Equation~\eqref{eq:V0_star_def_S} and \eqref{eq:Vm_star_def_S} show that the perturbed dynamics is identical to the original dynamics induced by Eqs.~\eqref{eq:Jump_V0_def_S} and \eqref{eq:Jump_Vc_def_S}, except that the end time is $\tau$ for the original dynamics, whereas $(1+\varepsilon)\tau = N \Delta t^\prime = N (1+\varepsilon)\Delta t$ for the perturbed dynamics. Therefore, from Eq.~\eqref{eq:C_scaling}, we obtain the scaling relation shown in Eqs.~\eqref{eq:mean_scaling_S} and \eqref{eq:var_scaling_S}.

\section{Comparison of lower bounds\label{sec:comparison}}
In the main text, we derived [Eq.~\mainUresultII]
\begin{equation}
\frac{\sqrbr{\mathcal{C}}^{2}}{\braket{\mathcal{C}}^{2}}\ge\frac{1}{\left|\mathrm{Tr}_{S}\left[e^{-iH_{\mathrm{eff}}\tau}\rho_{S}(0)\right]\right|^{-2}-1}.
\label{eq:main_result2_S}
\end{equation}
Meanwhile, in Ref.~\cite{Hasegawa:2020:TUROQS}, we showed the following similar result:
\begin{equation}
\frac{\sqrbr{\mathcal{C}}^{2}}{\braket{\mathcal{C}}^{2}}\ge\frac{1}{\mathrm{Tr}_{S}\left[e^{-iH_{\mathrm{eff}}^{\dagger}\tau}\rho_{S}(0)e^{iH_{\mathrm{eff}}\tau}\right]-1}.
\label{eq:OQ_TUR}
\end{equation}We now compare the tightness of Eqs.~\eqref{eq:main_result2_S} and \eqref{eq:OQ_TUR}.

As mentioned in the main text, the Lindblad equation [Eq.~\LindbladUdef{}] covers classical stochastic processes. 
A classical Markov chain with $N_S$ states can be represented quantum mechanically by an orthonormal basis
$\{\ket{b_1},\ket{b_2},...,\ket{b_{N_S}} \}$. 
Classical Markov chains can be emulated by setting $H_S = 0$, $L_{ji} = \sqrt{\gamma_{ji}}\ket{b_j}\bra{b_i}$, and
$\rho_S(t) = \sum_i p_i(t) \ket{b_i}\bra{b_i}$, where $\gamma_{ji}$ is a transition rate from $\ket{b_i}$ to $\ket{b_j}$, and $p_i(t)$ is the probability of being $\ket{b_i}$ at time $t$. 
The denominator of Eq.~\eqref{eq:main_result2_S} becomes
\begin{align}
\left|\mathrm{Tr}_{S}\left[e^{-iH_{\mathrm{eff}}\tau}\rho_{S}(0)\right]\right|^{-2}=\left|\mathrm{Tr}_{S}\left[e^{-\frac{\tau}{2}\sum_{m}L_{m}^{\dagger}L_{m}}\rho_{S}(0)\right]\right|^{-2}=\left(\sum_{i}e^{-\frac{\tau}{2}\sum_{j(\ne i)}\gamma_{ji}}p_{i}(0)\right)^{-2}.
\label{eq:main_result_classical}
\end{align}The denominator of Eq.~\eqref{eq:OQ_TUR} becomes
\begin{align}
\mathrm{Tr}_{S}\left[e^{-iH_{\mathrm{eff}}^{\dagger}\tau}\rho_{S}(0)e^{iH_{\mathrm{eff}}\tau}\right]&=\mathrm{Tr}_{S}\left[e^{\tau\sum_{m}L_{m}^{\dagger}L_{m}}\rho_{S}(0)\right]\nonumber\\&=\mathrm{Tr}_{S}\left[\sum_{i}e^{\tau\sum_{j(\ne i)}\gamma_{ji}}\ket{b_{i}}\bra{b_{i}}\sum_{j}p_{j}(0)\ket{b_{j}}\bra{b_{j}}\right]\nonumber\\&=\sum_{i}e^{\tau\sum_{j(\ne i)}\gamma_{ji}}p_{i}(0).
\label{eq:OQ_TUR_classical}
\end{align}According to Jensen's inequality, Eqs.~\eqref{eq:main_result_classical} and \eqref{eq:OQ_TUR_classical} satisfy the following inequality:
\begin{equation}
\left(\sum_{i}e^{-\frac{\tau}{2}\sum_{j(\ne i)}\gamma_{ji}}p_{i}\right)^{-2}\le\sum_{i}e^{\tau\sum_{j(\ne i)}\gamma_{ji}}p_{i},
\label{eq:TUR_compare_classical}
\end{equation}which indicates that Eq.~\eqref{eq:main_result2_S} yields a tighter lower bound than Eq.~\eqref{eq:OQ_TUR} does for the classical limit.

Next, we consider a quantum limit that $H_S \ne 0$ and $L_m = 0$.
Because the fidelity is always not larger than $1$, the denominator of Eq.~\eqref{eq:main_result2_S} is
\begin{align}
\left|\mathrm{Tr}_{S}\left[e^{-iH_{\mathrm{eff}}\tau}\rho_{S}(0)\right]\right|^{-2}=\left|\mathrm{Tr}_{S}\left[e^{-iH_{S}\tau}\rho_{S}(0)\right]\right|^{-2}\ge1.
\label{eq:LE_TUR_quantum}
\end{align}
The denominator of Eq.~\eqref{eq:OQ_TUR} becomes
\begin{align}
\mathrm{Tr}_{S}\left[e^{-iH_{\mathrm{eff}}^{\dagger}\tau}\rho_{S}(0)e^{iH_{\mathrm{eff}}\tau}\right]&=\mathrm{Tr}_{S}\left[\rho_{S}(0)\right]\nonumber\\&=1.
\label{eq:OQ_TUR_quantum}
\end{align}From Eqs.~\eqref{eq:LE_TUR_quantum} and \eqref{eq:OQ_TUR_quantum}, 
Eq.~\eqref{eq:OQ_TUR} yields a tighter lower bound than Eq.~\eqref{eq:main_result2_S} does for the quantum limit. 

In summary, when the system is close to a classical Markov process, the main result of this Letter provides a tighter bound.
Meanwhile, when the system is near a closed quantum system, the bound of Ref.~\cite{Hasegawa:2020:TUROQS} is tighter.

\section{Numerical simulation\label{sec:TUR_sim}}

\subsection{Model}

\begin{figure}
\centering
\includegraphics[width=12cm]{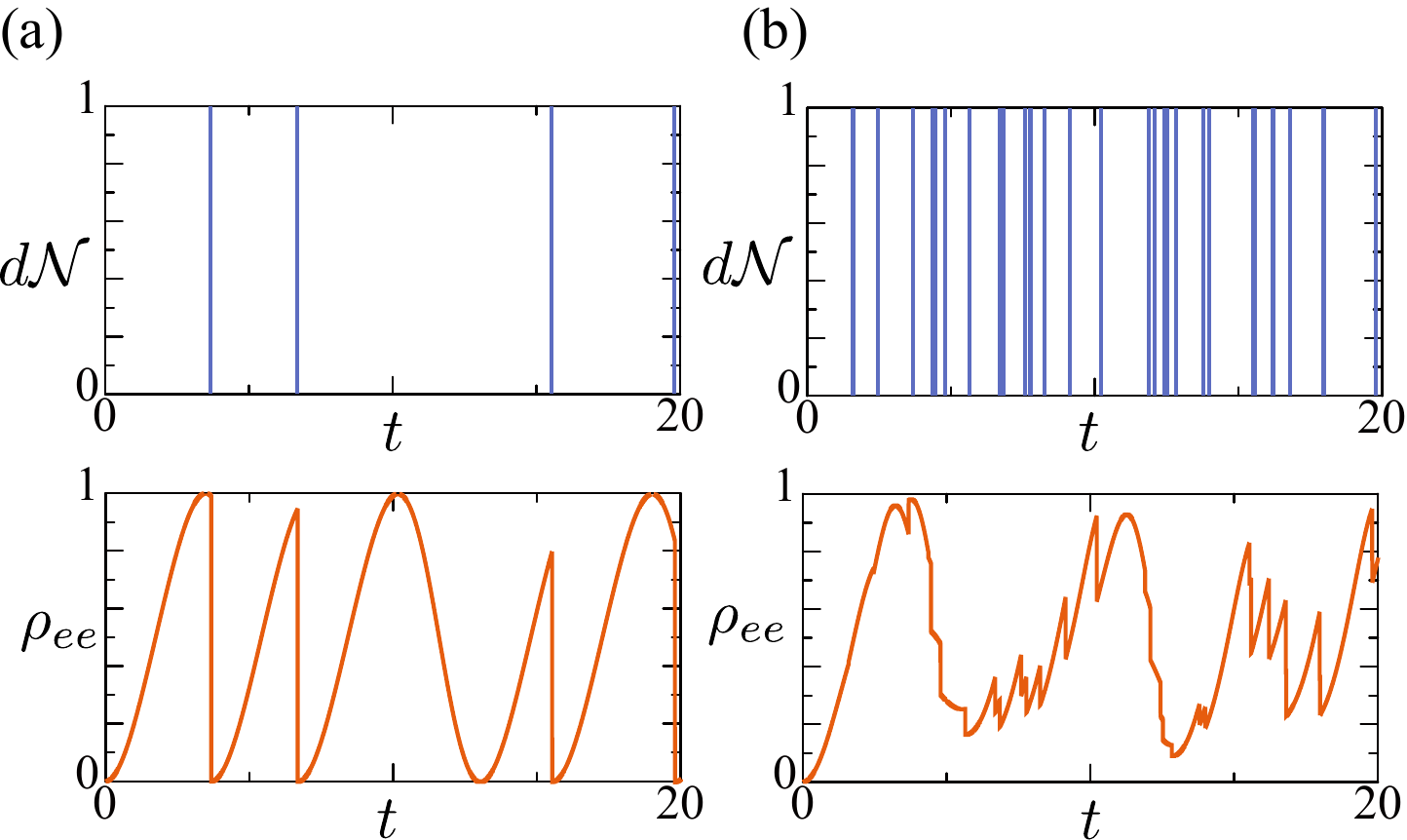} \caption{
Quantum trajectories generated by different measurement bases
for  a two-level atom driven by a laser field (see Section~\ref{sec:TUR_sim} for details).
Trajectories are generated by (a) $\zeta = 0$ (corresponding to the measurement basis $\{\ket{0},\ket{1}\}$) and (b) $\zeta = 1$. 
The upper panels describe the detection of jump events, and the lower panels show the population of the excited state
$\rho_{ee} \equiv \braket{\epsilon_e | \rho_S | \epsilon_e}$ as a function of $t$. 
\label{fig:quantum_traj}}
\end{figure}

We employ a two-level atom driven by a classical laser field. 
As stated in the main text, the Lindblad equation of the system is specified by
\begin{align}
H_{S}&=\Delta\ket{\epsilon_{e}}\bra{\epsilon_{e}}+\frac{\Omega}{2}(\ket{\epsilon_{e}}\bra{\epsilon_{g}}+\ket{\epsilon_{g}}\bra{\epsilon_{e}}),\label{eq:H_S_def}\\
L&=\sqrt{\kappa}\ket{\epsilon_g}\bra{\epsilon_e},\label{eq:L_atom_def}
\end{align}where $\ket{\epsilon_g}$ and $\ket{\epsilon_e}$ denote the ground and excited states, respectively;  $\Delta$, $\Omega$, and $\kappa$ are model parameters. 
The Kraus operators [Eqs.~\JumpUVOUdef{} and \JumpUVcUdef{}] are
\begin{align}
V_{0}&=\mathbb{I}_{S}-i\Delta tH_{S}-\frac{1}{2}\Delta tL^{\dagger}L,\label{eq:V0_sim_def}\\
V_{1}&=\sqrt{\Delta t}L,\label{eq:V1_sim_def}
\end{align}where $V_0$ and $V_1$ correspond to no detection and detection of a jump event within $\Delta t$, respectively. 
$V_0$ and $V_1$ correspond to the measurement basis of $\{\ket{0},\ket{1}\}$. 
Quantum trajectories generated by Eqs.~\eqref{eq:V0_sim_def} and \eqref{eq:V1_sim_def} can be described by the following 
stochastic Schr\"odinger equation:
\begin{equation}
d\rho_{S}=-i[H_{S},\rho_{S}]dt+\rho_{S}\mathrm{Tr}_{S}\left[L\rho_{S}L^{\dagger}\right]dt-\frac{\left\{ L^{\dagger}L,\rho_{S}\right\} }{2}dt+\left(\frac{L\rho_{S}L^{\dagger}}{\mathrm{Tr}_{S}[L\rho_{S}L^{\dagger}]}-\rho_{S}\right)d\mathcal{N},
\label{eq:SSE_sim_def}
\end{equation}where $d\mathcal{N}$ is a noise increment; $d\mathcal{N} = 1$ when a jump event (photon) is detected within $dt$, and is $0$ otherwise. 
The conditional expectation of $d\mathcal{N}$ is given by $\braket{d\mathcal{N}(t)}=\mathrm{Tr}_{S}[L\rho(t)L^{\dagger}]dt$, where $\rho(t)$ is a solution of 
Eq.~\eqref{eq:SSE_sim_def}.

As shown in Eq.~\eqref{eq:main_result_alpha}, the main result [Eq.~\mainUresultUTUR{}] holds for arbitrary measurements on $E$. 
As stated in the main text, 
the Lindblad equation is invariant under the following transformation:
\begin{align}
H_S &\to H_S - \frac{i}{2}(\zeta^* L - \zeta L^\dagger),\label{eq:H_transform}\\
L &\to L + \zeta \mathbb{I}_S,\label{eq:L_transform}
\end{align}where $\zeta \in \mathbb{C}$ can be an arbitrary value.  Substituting Eqs.~\eqref{eq:H_transform} and \eqref{eq:L_transform} into Eqs.~\eqref{eq:V0_sim_def} and \eqref{eq:V1_sim_def}, the corresponding Kraus operator is given by
\begin{align}
Y_{0}&=\mathbb{I}_{S}-i\Delta t\left[H_{S}-\frac{i}{2}(\zeta^{*}L-\zeta L^{\dagger})\right]-\frac{1}{2}\Delta t\left(L^{\dagger}+\zeta^{*}\right)\left(L+\zeta\right)\nonumber\\&=\mathbb{I}_{S}-i\Delta tH_{S}-\zeta^{*}L\Delta t-\frac{1}{2}\Delta t\left(L^{\dagger}L+|\zeta|^{2}\right),\label{eq:Y0_sim_def}\\Y_{1}&=\sqrt{\Delta t}\left(L+\zeta\mathbb{I}_{S}\right),\label{eq:Y1_sim_def}
\end{align}where $Y_0$ and $Y_1$ correspond to no detection and detection of a jump event, respectively. 
$Y_0$ and $Y_1$ [Eqs.~\eqref{eq:Y0_sim_def} and \eqref{eq:Y1_sim_def}] and $V_0$ and $V_1$ [Eqs.~\eqref{eq:V0_sim_def} and \eqref{eq:V1_sim_def}] are related through a unitary transformation shown in Eq.~\eqref{eq:Yn_def}, where $\mathcal{B}$ is given by
\begin{equation}
\mathcal{B}=\left[\begin{array}{cc}
-\frac{1}{2}|\zeta|^{2}\Delta t+1 & -\sqrt{\Delta t}\zeta^{*}\\
\sqrt{\Delta t}\zeta & -\frac{1}{2}|\zeta|^{2}\Delta t+1
\end{array}\right].\label{eq:unitary_B_def}
\end{equation}We have shown that the measurement basis is transformed through Eq.~\eqref{eq:phik_def},
where $\mathcal{A} = \mathcal{B}^\dagger$. 
Since Eq.~\eqref{eq:unitary_B_def} shows that we can specify $\mathcal{B}$ through $\zeta$ alone, 
we can change the measurement basis by changing $\zeta$. 
Figure~\ref{fig:quantum_traj} shows trajectories 
generated by (a) $\zeta = 0$ and (b) $\zeta = 1$.
As explained above, the trajectories of  Fig.~\ref{fig:quantum_traj}(a) and (b) use different measurement bases. 
Although the trajectories of Fig.~\ref{fig:quantum_traj}(a) and (b) are different, 
both cases reduce to the same dynamics on average.

\begin{figure}
\centering
\includegraphics[width=13cm]{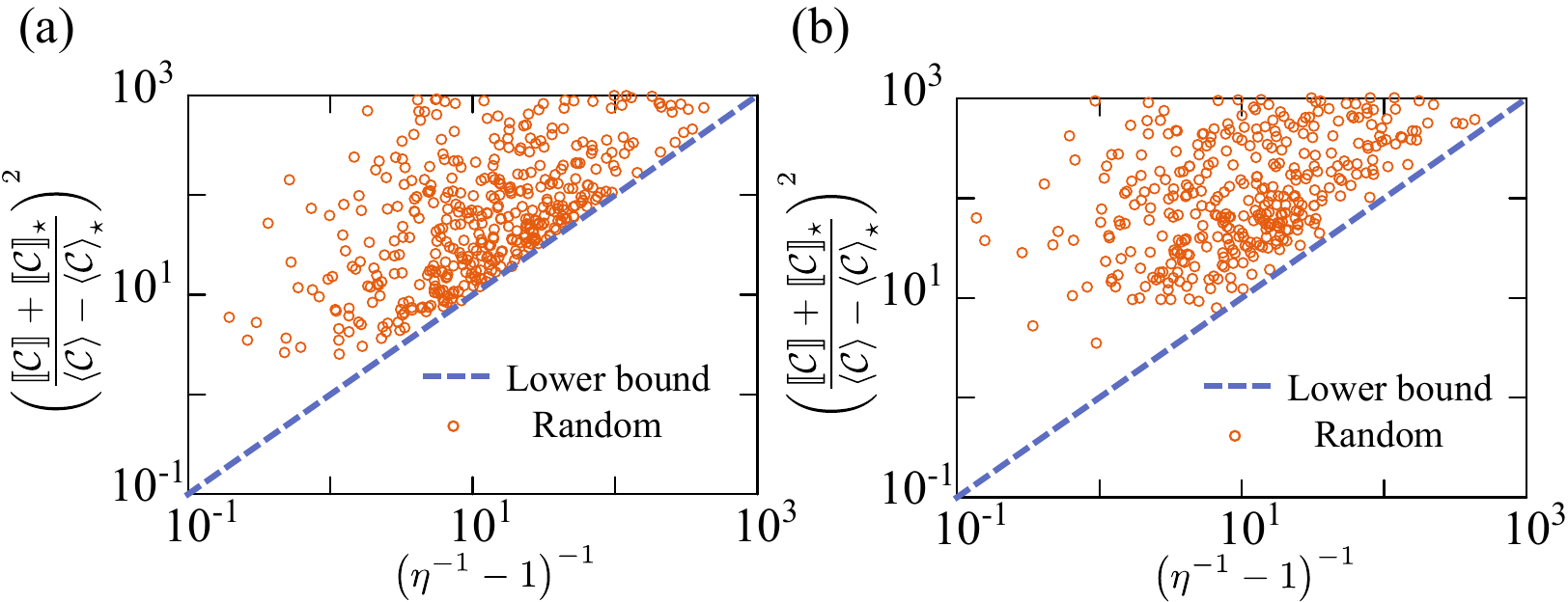} \caption{
Precision $(\sqrbr{\mathcal{C}}+\sqrbr{\mathcal{C}}_{\star})^{2}/(\braket{\mathcal{C}}-\braket{\mathcal{C}}_{\star})^{2}$
as a function of $(\eta^{-1} - 1)^{-1}$ for random realizations, where $\eta = |\mathrm{Tr}_S[e^{\mathcal{K}\tau}\rho_S(0)]|^2$. 
Random realizations are plotted by circles and the lower bound by the dashed lines. 
Random realizations are calculated with (a) fixed continuous measurement $V_m$ and (b) random continuous measurement $Y_m$ by randomly sampling $\zeta$ from $|\zeta| \in [0.0, 1.0]$.
In (a) and (b), the parameter ranges are $\Delta \in [0.1, 3.0]$, $\Omega \in [0.1,3.0]$, and $\kappa \in [0.1,3.0]$ for
each dynamics, and $\tau \in [0.1, 1.0]$. 
\label{fig:simulation_supp}}
\end{figure}

\subsection{Simulation of Eq.~\mainUresultUTUR{}}

We conduct a numerical simulation to verify Eq.~\mainUresultUTUR{}.
Equation~\mainUresultUTUR{} considers two dynamics, the original and its perturbed dynamics. For each dynamics, 
we randomly select the model parameters $\Delta$, $\Omega$, and $\kappa$ (i.e., six model parameters in total).
The time $\tau$ is randomly sampled
(see the caption of Fig.~\ref{fig:simulation_supp} for the parameter ranges), and 
the initial density operator $\rho_S$ is randomly determined. 
For the selected parameters and density operator, we generate different quantum trajectories and calculate $(\sqrbr{\mathcal{C}}+\sqrbr{\mathcal{C}}_{\star})^{2}/(\braket{\mathcal{C}}-\braket{\mathcal{C}}_{\star})^{2}$. 
In Fig.~\ref{fig:simulation_supp}(a), we plot $(\sqrbr{\mathcal{C}}+\sqrbr{\mathcal{C}}_{\star})^{2}/(\braket{\mathcal{C}}-\braket{\mathcal{C}}_{\star})^{2}$ as a function of $(\eta^{-1}-1)^{-1}$ with $\eta=|\mathrm{Tr}_{S}[e^{\mathcal{K}\tau}\rho_S(0)]|^{2}$ (which is the right-hand side of Eq.~\mainUresultUTUR{}) by circles, where the dashed line denotes the lower bound. 
All circles are located above the line, verifying Eq.~\mainUresultUTUR{} for the driven two-level atom system.

As denoted above, Eq.~\mainUresultUTUR{} should hold for any continuous measurement (any unraveling). 
Besides, we conduct a numerical simulation for different continuous measurements.
First, we randomly determine $\zeta$ to change the continuous measurement.
Then, we follow the same procedure as in Fig.~\ref{fig:simulation_supp}(a). 
In Fig.~\ref{fig:simulation_supp}(b), we plot $(\sqrbr{\mathcal{C}}+\sqrbr{\mathcal{C}}_{\star})^{2}/(\braket{\mathcal{C}}-\braket{\mathcal{C}}_{\star})^{2}$ as a function of $(\eta^{-1}-1)^{-1}$ with $\eta=|\mathrm{Tr}_{S}[e^{\mathcal{K}\tau}\rho_S(0)]|^{2}$ by circles, where the dashed line denotes the lower bound of Eq.~\mainUresultUTUR{}. 
All circles are above the line, verifying Eq.~\mainUresultUTUR{} for
different continuous measurements.

\section{Derivation of lower bound of Hellinger distance\label{sec:Hellinger_bound_derivation}}

For convenience, we briefly explain the derivation of the lower bound of the Hellinger distance in Ref.~\cite{Nishiyama:2020:HellingerBound}. For the detailed derivation, see the original paper \cite{Nishiyama:2020:HellingerBound}.

Let $P(x)$ and $Q(x)$ be probability distributions defined on a discrete set $x \in \{x_1,x_2,\ldots,x_\mathfrak{N}\}$, where $x_i \in \mathbb{R}$, $x_i \ne x_j$ for $i \ne j$, and 
$\mathfrak{N}$ is the number of elements in the set. 
The Hellinger distance between $P(x)$ and $Q(x)$ is defined by
\begin{equation}
\mathcal{H}^{2}(P,Q)=\frac{1}{2}\sum_{x\in\{x_{1},x_{2},\ldots,x_{\mathfrak{N}}\}}\left(\sqrt{P(x)}-\sqrt{Q(x)}\right)^{2}=1-\sum_{x\in\{x_{1},x_{2},\ldots,x_{\mathfrak{N}}\}}\sqrt{P(x)Q(x)}.
\label{eq:Hellinger_def_S}
\end{equation}We will obtain the lower bound for the Hellinger distance, given the means and variances of $P(x)$ and $Q(x)$. 
Let $\mu_P$ and $\mu_Q$ be the means of $P(x)$ and $Q(x)$, respectively, and $\sigma_P^2$ and $\sigma_Q^2$ be variances of $P(x)$ and $Q(x)$, respectively. 

We first consider a case of $\mathfrak{N}=2$, i.e., $P(x)$ and $Q(x)$ are defined on only two points $\{x_1,x_2\}$. 
Given the means ($\mu_P$ and $\mu_Q$) and variances ($\sigma_P^2$ and $\sigma_Q^2$),
$(x_1,x_2,\mathfrak{p}_1,\mathfrak{p}_2,\mathfrak{q}_1,\mathfrak{q}_2)$, where $\mathfrak{p}_1 \equiv P(x_1)$, $\mathfrak{p}_2\equiv  P(x_2)$, $\mathfrak{q}_1 \equiv Q(x_1)$, and $\mathfrak{q}_2\equiv Q(x_2)$, can be obtained by solving the following algebraic equation:
\begin{align}
\mu_{P}&=x_{1}\mathfrak{p}_{1}+x_{2}\mathfrak{p}_{2},\;\;\;\sigma_{P}^{2}=x_{1}^{2}\mathfrak{p}_{1}+x_{2}^{2}\mathfrak{p}_{2}-\mu_{P}^{2},\;\;\;1=\mathfrak{p}_{1}+\mathfrak{p}_{2},\label{eq:Pcondition}\\
\mu_{Q}&=x_{1}\mathfrak{q}_{1}+x_{2}\mathfrak{q}_{2},\;\;\;\sigma_{Q}^{2}=x_{1}^{2}\mathfrak{q}_{1}+x_{2}^{2}\mathfrak{q}_{2}-\mu_{Q}^{2},\;\;\;1=\mathfrak{q}_{1}+\mathfrak{q}_{2}.\label{eq:Qcondition}
\end{align}The solution $(x_1,x_2,\mathfrak{p}_1,\mathfrak{p}_2,\mathfrak{q}_1,\mathfrak{q}_2)$ of Eqs.~\eqref{eq:Pcondition} and \eqref{eq:Qcondition}, which is unique, except for the apparent symmetry ($x_{1}\leftrightarrow x_2$, $\mathfrak{p}_1 \leftrightarrow \mathfrak{p}_2$, and $\mathfrak{q}_1 \leftrightarrow \mathfrak{q}_2$), is given by
\begin{equation}
x_{1}=\mu_{P}+\sqrt{\frac{\mathfrak{p}_{2}\sigma_{P}^{2}}{\mathfrak{p}_{1}}},\,\,\,x_{2}=\mu_{P}-\sqrt{\frac{\mathfrak{p}_{1}\sigma_{P}^{2}}{\mathfrak{p}_{2}}},\,\,\,\mathfrak{p}_{1}=\frac{1}{2}+\frac{\mathfrak{b}+\mathfrak{a}^{2}}{4\mathfrak{a}\mathfrak{c}},\,\,\,\mathfrak{p}_{2}=1-\mathfrak{p}_{1},\,\,\,\mathfrak{q}_{1}=\frac{1}{2}+\frac{\mathfrak{b}-\mathfrak{a}^{2}}{4\mathfrak{a}\mathfrak{c}},\,\,\,\mathfrak{q}_{2}=1-\mathfrak{q}_{1},
\label{eq:Hellinger_solution_S}
\end{equation}where $\mathfrak{a}$, $\mathfrak{b}$, and $\mathfrak{c}$ are defined by 
\begin{equation}
\mathfrak{a}\equiv\mu_{P}-\mu_{Q},\,\,\,\mathfrak{b}\equiv\sigma_{Q}^{2}-\sigma_{P}^{2},\,\,\,\mathfrak{c}\equiv\frac{1}{2|\mathfrak{a}|}\sqrt{\mathfrak{a}^{4}+\mathfrak{b}^{2}+2\mathfrak{a}^{2}(\sigma_{P}^{2}+\sigma_{Q}^{2})}.
\label{eq:abc_def}
\end{equation}Substituting the solution of Eq.~\eqref{eq:Hellinger_solution_S} into Eq.~\eqref{eq:Hellinger_def_S}, the Hellinger distance for $\mathfrak{N}=2$ becomes
\begin{equation}
\mathcal{H}^2(P,Q) = \mathfrak{h}^2(\mu_P,\mu_Q,\sigma_P,\sigma_Q),
\label{eq:H2_def}
\end{equation}where
\begin{equation}
\mathfrak{h}^{2}(\mu_{P},\mu_{Q},\sigma_{P},\sigma_{Q})\equiv1-\left(1+\frac{(\mu_{P}-\mu_{Q})^{2}}{(\sigma_{P}+\sigma_{Q})^{2}}\right)^{-\frac{1}{2}}.
\label{eq:h2_def}
\end{equation}Therefore, for $\mathfrak{N}=2$, the Hellinger distance is uniquely specified, given the means ($\mu_P$ and $\mu_Q$) and variances ($\sigma_P^2$ and $\sigma_Q^2$). 

Next, we consider a case of $\mathfrak{N} > 2$. We define auxiliary variables $\mathfrak{r}_i \in \mathbb{R}$ and $\mathfrak{s}_i \in \mathbb{R}$, which satisfy
\begin{equation}
P(x_i) = \mathfrak{r}_i^2,\;\;\;Q(x_i) = \mathfrak{s}^2_i,\;\;\;(i = 1,2,\ldots,\mathfrak{N}).
\end{equation}Here, owing to the normalization of $P(x)$ and $Q(x)$, $\mathfrak{r}_i$ and $\mathfrak{s}_i$ should satisfy
\begin{equation}
\sum_{i=1}^\mathfrak{N} \mathfrak{r}_i^2 =1,\;\;\;\sum_{i=1}^\mathfrak{N} \mathfrak{s}_i^2 =1.
\label{eq:rs_condition}
\end{equation}Given the means ($\mu_P$ and $\mu_Q$) and variances ($\sigma_P^2$ and $\sigma_Q^2$), $\{x_i\}_{i=1}^\mathfrak{N}$, $\{\mathfrak{r}_i\}_{i=1}^\mathfrak{N}$, and $\{\mathfrak{s}_i\}_{i=1}^\mathfrak{N}$ should satisfy the following constraint:
\begin{equation}
\sum_{i=1}^{\mathfrak{N}}x_{i}\mathfrak{r}_{i}^{2}-\mu_{P}=0,\,\,\,\sum_{i=1}^{\mathfrak{N}}x_{i}\mathfrak{s}_{i}^{2}-\mu_{Q}=0,\,\,\,\sum_{i=1}^{\mathfrak{N}}x_{i}^{2}\mathfrak{r}_{i}^{2}-\mu_{P}^{2}-\sigma_{P}^{2}=0,\,\,\,\sum_{i=1}^{\mathfrak{N}}x_{i}^{2}\mathfrak{s}_{i}^{2}-\mu_{Q}^{2}-\sigma_{Q}^{2}=0.
\label{eq:constraint_S}
\end{equation}According to the second expression of the Hellinger distance in Eq.~\eqref{eq:Hellinger_def_S}, minimizing the Hellinger distance is identical to maximizing $\sum_{i=1}^\mathfrak{N} \mathfrak{r}_i \mathfrak{s}_i$ under the constraints of Eqs.~\eqref{eq:rs_condition} and \eqref{eq:constraint_S}. 
This constrained optimization can be solved through the method of Lagrange multiplier. 
Let us introduce the following Lagrangian:
\begin{align}
\mathfrak{L}&\equiv\sum_{i=1}^{\mathfrak{N}}\mathfrak{r}_{i}\mathfrak{s}_{i}+\lambda_{1}\left(\sum_{i=1}^{\mathfrak{N}}\mathfrak{r}_{i}^{2}-1\right)+\lambda_{2}\left(\sum_{i=1}^{\mathfrak{N}}\mathfrak{s}_{i}^{2}-1\right)+\lambda_{3}\left(\sum_{i=1}^{\mathfrak{N}}x_{i}\mathfrak{r}_{i}^{2}-\mu_{P}\right)+\lambda_{4}\left(\sum_{i=1}^{\mathfrak{N}}x_{i}\mathfrak{s}_{i}^{2}-\mu_{Q}\right)\nonumber\\&+\lambda_{5}\left(\sum_{i=1}^{\mathfrak{N}}x_{i}^{2}\mathfrak{r}_{i}^{2}-\mu_{P}^{2}-\sigma_{P}^{2}\right)+\lambda_{6}\left(\sum_{i=1}^{\mathfrak{N}}x_{i}^{2}\mathfrak{s}_{i}^{2}-\mu_{Q}^{2}-\sigma_{Q}^{2}\right),
\label{eq:Lagrangian_def}
\end{align}where $\lambda_i$ are Lagrange multipliers for the constraints of Eqs.~\eqref{eq:rs_condition} and \eqref{eq:constraint_S}. Optimal solutions $\{x_i\}_{i=1}^\mathfrak{N}$, $\{\mathfrak{r}_i\}_{i=1}^\mathfrak{N}$, and $\{\mathfrak{s}_i\}_{i=1}^\mathfrak{N}$, which give the minimum of the Hellinger distance, can be obtained by solving
\begin{align}
\frac{\partial\mathfrak{L}}{\partial\mathfrak{r}_{i}}&=\mathfrak{s}_{i}+\mathfrak{r}_{i}\left(2\lambda_{1}+2\lambda_{3}x_{i}+2\lambda_{5}x_{i}^{2}\right)=0,\label{eq:LM1}\\
\frac{\partial\mathfrak{L}}{\partial\mathfrak{s}_{i}}&=\mathfrak{r}_{i}+\mathfrak{s}_{i}\left(2\lambda_{2}+2\lambda_{4}x_{i}+2\lambda_{6}x_{i}^{2}\right)=0,\label{eq:LM2}\\
\frac{\partial\mathfrak{L}}{\partial x_{i}}&=\lambda_{3}\mathfrak{r}_{i}^{2}+\lambda_{4}\mathfrak{s}_{i}^{2}+2\lambda_{5}x_{i}\mathfrak{r}_{i}^{2}+2\lambda_{6}x_{i}\mathfrak{s}_{i}^{2}=0,\label{eq:LM3}
\end{align}for all $i = 1,2,\ldots,\mathfrak{N}$. Combining Eqs.~\eqref{eq:LM1} and \eqref{eq:LM2} yields
\begin{equation}
\mathfrak{F}(x_i)\equiv\left(2\lambda_{1}+2\lambda_{3}x_{i}+2\lambda_{5}x_{i}^{2}\right)\left(2\lambda_{2}+2\lambda_{4}x_{i}+2\lambda_{6}x_{i}^{2}\right)-1=0\;\;\;(i=1,2,\ldots,\mathfrak{N}).
\label{eq:frakF_def}
\end{equation}Moreover, by subtracting
 a term Eqs.~\eqref{eq:LM1} multiplied by $\mathfrak{r}_i$ from a term \eqref{eq:LM2} multiplied by $\mathfrak{s}_i$, and substituting Eq.~\eqref{eq:LM3} into the resulting equation, the derivative of $\mathfrak{F}(x_i)$ vanishes:
\begin{equation}
\frac{d}{dx_{i}}\mathfrak{F}(x_{i})=16x^{3}\lambda_{5}\lambda_{6}+12x^{2}\lambda_{3}\lambda_{6}+12x^{2}\lambda_{4}\lambda_{5}+8x\lambda_{1}\lambda_{6}+8x\lambda_{2}\lambda_{5}+8x\lambda_{3}\lambda_{4}+4\lambda_{1}\lambda_{4}+4\lambda_{2}\lambda_{3}=0\;\;\;(i=1,2,\ldots,\mathfrak{N}).
\label{eq:dFdxi}
\end{equation}From Eq.~\eqref{eq:frakF_def}, there are at most four solutions with respect to $x_i$ in $\mathfrak{F}(x_i) = 0$ since it is a quartic function of $x_i$. However, owing to Eq.~\eqref{eq:dFdxi}, each solution degenerates, indicating that there are at most two distinct solutions in $\mathfrak{F}(x_i) = 0$. When there is only one solution, the variance vanishes. Therefore, when considering only $\sigma_P >0$ and $\sigma_Q>0$, $\mathfrak{F}(x_i) = 0$ has two distinct solutions (vanishing variance cases were handled in the original paper \cite{Nishiyama:2020:HellingerBound}). This result indicates that, when we minimize the Hellinger distance, given the means ($\mu_P$ and $\mu_Q$) and variances ($\sigma_P^2$ and $\sigma_Q^2$), the case of $\mathfrak{N}=2$ gives the lower bound for the Hellinger distance. 
Because the case of $\mathfrak{N}=2$ has been derived in Eq.~\eqref{eq:H2_def}, the following relation gives the lower bound for the Hellinger distance for arbitrary $\mathfrak{N}\ge 2$:
\begin{equation}
\mathcal{H}^2(P,Q) \ge \mathfrak{h}^2(\mu_P,\mu_Q,\sigma_P,\sigma_Q).
\label{eq:H2_ineq}
\end{equation}Moreover, the equality of Eqs.~\eqref{eq:H2_ineq} and \eqref{eq:h2_def} is attained if and only if $P(x)$ and $Q(x)$ are defined on two points ($\mathfrak{N}=2$).